# Incremental Bidirectional Typing via Order Maintenance

THOMAS J. PORTER, University of Michigan, USA
MARISA KIRISAME, University of Utah, USA
IVAN WEI, University of Michigan, USA
PAVEL PANCHEKHA, University of Utah, USA
CYRUS OMAR, University of Michigan, USA

*Live programming environments* provide various semantic services, including type checking and evaluation, continuously as the user is editing the program. The live paradigm promises to improve the developer experience, but liveness is an implementation challenge particularly when working with large programs. This paper specifies and efficiently implements a system the is able to incrementally update type information for a live program in response to fine-grained program edits. This information includes type error marks and information about the expected and actual type on every expression. The system is specified type-theoretically as a small-step dynamics that propagates updates through the marked and annotated program. Most updates flow according to a base bidirectional type system. Additional pointers are maintained to connect bound variables to their binding locations, with type updates traversing these pointers directly. Order maintenance data structures are employed to efficiently maintain these pointers and to prioritize the order of update propagation. We prove this system is equivalent to naive reanalysis in the Agda theorem prover, along with other important metatheoretic properties. We then provide an efficient OCaml implementation, detailing a number of impactful optimizations. We evaluate this implementation's performance with a large stress-test and find that it is able to achieve dramatic speed-ups of $275.96\times$ compared to from-scratch reanalysis.

## 1 Introduction

Modern programming environments increasingly aim to provide *live* (i.e. continuously available) semantic feedback to the programmer throughout the editing process, even when the program is incomplete or contains localized errors [24]. Traditional programming language implementations are designed for batch usage and expect complete programs, so they often struggle to keep up with the demands of live programming environments.

This paper focuses on the problem of providing live *type information* while a program sketch (i.e. a program with syntactic *holes* in various positions) is being edited. By type information, we mean the locations and causes of type errors as well as information about the expected and actual type at *every* location in the program sketch, even in the presence of type errors at other locations. In order to achieve liveness, our central constraint is that the system must not need to traverse the entire program sketch (which we assume to be arbitrarily large) between edits. Instead, the system should be able to *incrementally* update the collected type information, at a computational cost proportional to the number of potentially affected locations.

Even this essential computational cost may be inherently high for some changes to large programs, so we go further and aim to minimize situations where editing is blocked waiting for updates to the type information to propagate, and eliminate situations where updates caused by previous edits are rolled back and recomputed when a subsequent edit is performed. Instead, we allow updates to propagate at finite speed through the program sketch, even as new edits come in, with correctness guaranteed once this interleaving of edits and updates has fully quiesced.

These uncompromising technical aims are motivated both by the increasing prevalence of multi-million-line code bases (often organized into "monorepos") in many organizations, and by a vision of the future of programming in which large-scale scientific and social collaborations occur within a shared, live programming environment that allows thousands of participants to collaboratively edit a single "planetary-scale" live program [5, 16].

In pursuit of these aims, this paper develops a foundational type-theoretic calculus of incremental type information maintenance for a bidirectional type system (i.e. a type system organized around local type inference) [11, 26]. We equip this calculus with a comprehensive metatheory mechanized in the Agda proof assistant. Based on this core calculus, we develop an implementation that employs techniques pioneered in web browsers for incremental page layout computations [17] —most centrally, the use of order maintenance data structures [6]—to realize the promised incremental speedups.

We start in Section 2 by reviewing the necessary background on bidirectional typing [11, 26], gradual typing [30, 31], and the *marked lambda calculus (MLC)* [37], which specifies a total procedure for bidirectional type error marking. In Section 3 we develop a variant of this calculus, called the *marked and annotated lambda calculus (MALC)*, which simplifies the specification of error marking and adds additional annotations recording the expected and inferred types at each location in the program. Collectively, we call error marks and type annotations *type information*. We also introduce a calculus of purely syntactic structural edit actions on MALC terms. We expect that these tree-structured edit actions will be generated either directly by a structure editor like Hazel [23], Scratch [19], or Pantograph [27], or else that they will be inferred by a parser together with a tree differencing algorithm [8]; incrementalizing these systems is beyond the scope of this paper.

We then proceed to the central problem confronted by this paper: incrementalizing the static semantics of MALC in response to these structural edits. We start with an overview of Incremental MALC by example in Section 4 and then fully specify its semantics and metatheory in Section 5. The key idea is to specify the semantics of type information updates as a small-step *update propagation* dynamics. During update propagation, the calculus maintains an *update propagation frontier* through the use of dirty bits on the types within MALC terms. Each update propagation step considers a dirty type at the frontier, calculates its local ramifications, and propagates the frontier correspondingly. This process is kicked off by Incremental MALC's edit action semantics, which creates initial dirty bits at the location of the edit, and critically, at other locations implied by the binding structure of the program. We show that edit actions and updates can be interleaved confluently. Consequently, editing is only blocked for the duration of individual edits and propagation steps, which are generally very short. We establish that correctness (with respect to MALC's semantics) is ensured once update propagation quiesces, i.e. when the frontier is empty.

Incremental MALC specifies certain critical operations related to bindings declaratively, in the usual type-theoretic style, and naive implementations of these operations would require substantial subtree traversals. We turn in Section 6 to developing the data structures and algorithms necessary to efficiently implement Incremental MALC. We call our implementation Malcom, a portmanteau of MALC and "order maintenance" due to the central role of efficient order maintenance data structures in the implementation. We then evaluate the performance of Malcom in Section 7, comparing the incremental implementation to the from-scratch implementation. We find that for partially randomized edits to a large synthetic functional program designed to serve as a stress-test, a $275.96\times$ speedup is obtained relative to the same system in from-scratch mode. We conclude with a review of related work in Section 8 and discussion and directions for future research in Section 9.

## 2 Background

The type system considered in this paper is a variant of the marked lambda calculus (MLC), introduced recently by Zhao et al. [37]. The marked lambda calculus combines **bidirectional typing** and **gradual typing** to achieve "total type error localization and recovery," that is, the ability to localize type errors occurring throughout a program, without the presence of one error preventing the localization of another. The key components of this approach are summarized below.

*Bidirectional Typing*    Bidirectional typing is an approach to type checking that splits the ordinary typing judgement into two judgements: an analytic judgment $\Gamma \vdash e \Leftarrow \tau$ for checking whether the expression $e$ has type $\tau$ under context $\Gamma$, and a synthetic judgment $\Gamma \vdash e \Rightarrow \tau$, which infers a type from $e$ [11, 26]. The inference rules for the analytic and synthetic judgments directly specify a typing algorithm, and because of the locality of the flow of typing information, errors are naturally localized to the AST node where analysis or synthesis fails, unlike with unification-based inference. Zhao et al. [37] also give a unification-based type inference system on top of the bidirectionally typed core, but we do not consider this extension in this paper.

*Gradual Typing*    Gradual typing allows type checking intermixed typed and untyped code. The gradually typed lambda calculus extends the simply typed lambda calculus with the unknown (or dynamic) type, ?, and replaces type equality with a type consistency relation [30, 31]. The unknown type is consistent with every type. In live programming environments, gradual typing allows the type checker to handle type-incorrect programs that occur as the user edits.

*Error Marks*    The marked lambda calculus centers around the *marking* procedure, a total function that transforms an ordinary expression $e$ into a marked expression $\check{e}$ by inserting error marks. Error marks can be attached to an expression to indicate the presence and nature of a type error localized to that expression. Intuitively, error marks are formal representations of the "red squiggles" one might see in a program editor.

The marking procedure is defined bidirectionally, with an analytic marking judgment $\Gamma \vdash e \rightsquigarrow \check{e} \Leftarrow \tau$ and a synthetic marking judgment $\Gamma \vdash e \rightsquigarrow \check{e} \Rightarrow \tau$. These have the same interpretations as the ordinary bidirectional judgments, but with the additional output $\check{e}$ (pronounced "$e$ check") and the additional property of specifying total functions. Any input program $e$, no matter how ill-typed, can be marked in either mode. The key to this totality is the use of the gradual type to recover from localized errors. Where a type error would occur in a simple type system, a term is assigned the unknown type in the marked lambda calculus. This allows typing to proceed optimistically, as the unknown type is consistent with all types and will not introduce additional downstream errors. The following example shows the case of a number literal applied to a free variable; the number is marked for synthesizing a non-function type while in function position, the variable is marked as free, and the program synthesizes the unknown type. Here, $(\!|e|\!)$ is a marked expression and the subscripts and superscripts are simply symbolic representations of the "error message".

$$\varnothing \vdash 1 \; x \rightsquigarrow (\!|1|\!)^{\overrightarrow{\blacktriangleright}}_{\blacktriangleright_{+}} \; (\!|x|\!)_{\square} \Rightarrow \text{?}$$

## 3   The Marked and Annotated Lambda Calculus

We now introduce the marked and annotated lambda calculus (MALC), which modifies MLC in two ways in support of the goals of this paper. First, in MALC, terms store boolean marks indicating whether an error has been localized to that term, rather than marks being represented as term constructors. Second, every term is annotated with optional analyzed and synthesized types.

### 3.1   Syntax

The syntax of MALC is given in Figure 8. Types, $\tau$, are drawn from a standard simple type system, with the addition of the unknown type of gradual type theory, ?, which also serves as a type hole. We omit base types for simplicity. Bare expressions, $e$, are the ordinary expression terms in a simply typed language, consisting of variables, function abstractions, function applications, and type ascriptions. Our notation is nonstandard, using $\mapsto$ in abstractions and $\triangleleft$ in applications; this is to provide a syntactic anchor for marks on these forms. Bare expressions also allow for expression holes, $?$, representing the incomplete parts of a program that occur as a user edits a program. We likewise allow the binding variable, $x?$, in a lambda abstraction to be a "binding hole," which binds nothing, to support flexible editing.

| $x$ | $\in$ | Variable | | |
|-----|-------|----------|------|---|
| $x?$ | $\in$ | Binding | ::= | $? \mid x$ |
| $\tau$ | $\in$ | Type | ::= | $? \mid \tau \to \tau$ |
| $e$ | $\in$ | BareExp | ::= | $? \mid x \mid \lambda x? : \tau \mapsto e \mid e \triangleright e \mid e : \tau$ |
| $m$ | $\in$ | Mark | ::= | $\checkmark \mid \times$ |
| $\sigma$ | $\in$ | TypeOpt | ::= | $\square \mid \tau$ |
| $\check{e}^{\Rightarrow}$ | $\in$ | MarkedSynExp | ::= | $\dot{\check{e}} \overset{\sigma}{\Rightarrow}$ |
| $\dot{\check{e}}$ | $\in$ | MarkedConExp | ::= | $? \mid x_m \mid \lambda x? : \tau \mapsto_{m,m} {}^{\Rightarrow}\check{e} \mid {}^{\Rightarrow}\check{e} \triangleleft_m {}^{\Rightarrow}\check{e} \mid {}^{\Rightarrow}\check{e} : \tau$ |
| ${}^{\Rightarrow}\check{e}$ | $\in$ | MarkedAnaExp | ::= | $\overset{\sigma}{\Rightarrow}_m \check{e}^{\Rightarrow}$ |
| $\check{p}$ | $\in$ | MarkedProgram $\subseteq$ MarkedAnaExp | ::= | $\overset{\square}{\Rightarrow}_{\checkmark} \check{e}^{\Rightarrow}$ |

Fig. 1. Syntax of MALC

Marked expressions are like bare expressions but with additional type information. Marked expressions are defined by three mutually recursive sorts: marked synthetic expressions, $\check{e}^{\Rightarrow}$, marked constructor expressions, $\dot{\check{e}}$, and marked analytic expressions, ${}^{\Rightarrow}\check{e}$. The marked synthetic expression constructor stores an optional synthesized type, $\sigma$; the marked constructor expression constructors are the core syntactic forms; and the marked analytic expression constructor stores an optional analyzed type. The arrows suggest a flow of information from left to right: analyzed types come from the left, and synthesized types proceed to the right. Marked analytic expressions also include a consistency mark, $m$, which indicates consistency between the analyzed and synthesized types: $\times$ indicates the presence of a error and $\checkmark$ indicates the absence of one.

Marked programs, $\check{p}$, are modeled as analytic expressions with no analyzed type. They are not just synthetic expressions because the incremental version of the calculus makes use of the empty analytic data at the root of the program.

In addition to the consistency marks, marked constructor expressions also have marks. Each mark position corresponds to a possible kind of static error. Variables have a mark indicating whether they are free. Function abstractions have two marks, the first corresponding to whether the abstraction is analyzed against a non-function type, and the second corresponding to whether the domain of the analyzed type is inconsistent with the annotation on the abstraction. Function applications have a mark indicating whether the first child synthesizes a non-function type.

The formal syntax is quite elaborate, but most of it would be hidden from the user by default, with error marks only displayed if set to $\times$ and local types only displayed when queried.

## 3.2 Marking

Marking in MALC is given by the mutually recursive synthetic and analytic marking judgments, defined in Figure 10. Note that the synthetic marking judgment $\Gamma \vdash e \rightsquigarrow \check{e}^{\Rightarrow}$ does not explicitly output the synthesized type because it can be found in the type annotation on $\check{e}^{\Rightarrow}$. Also note that the analytic judgment $\Gamma \vdash \tau \Rightarrow e \rightsquigarrow {}^{\Rightarrow}\check{e}$ is written with the type on the left instead of the right; we adopt this convention so that information tends to flow from left to right.

The MARKHOLE rule marks a bare expression hole, $?$, as an expression hole synthesizing the unknown type.

$$\boxed{\Gamma \vdash e \leadsto \check{e}^{\Rightarrow}}$$

**MarkHole**

$$\Gamma \vdash ? \leadsto ?^{\,?}_{\Rightarrow}$$

**MarkVar**

$$\frac{x_m : \tau \in \Gamma}{\Gamma \vdash x \leadsto x_m{}^{\tau}_{\Rightarrow}}$$

**MarkAsc**

$$\frac{\Gamma \vdash \tau \Rightarrow e \leadsto {}^{\Rightarrow}\check{e}}{\Gamma \vdash e : \tau \leadsto {}^{\Rightarrow}\check{e} : \tau {}^{\tau}_{\Rightarrow}}$$

**MarkSynFun**

$$\frac{\Gamma,\, x? : \tau_1 \vdash e \leadsto \dot{\check{e}} {}^{\tau_2}_{\Rightarrow}}{\Gamma \vdash \lambda x? : \tau_1 \mapsto e \leadsto \lambda x? : \tau_1 \mapsto_{\checkmark,\checkmark} \left( {}^{\Box}_{\Rightarrow}{}_{\checkmark} \dot{\check{e}} {}^{\tau_2}_{\Rightarrow} \right) {}^{\tau_1 \to \tau_2}_{\Rightarrow}}$$

**MarkAp**

$$\frac{\Gamma \vdash e_1 \leadsto \dot{\check{e}} {}^{\tau}_{\Rightarrow} \qquad \tau \blacktriangleright^{\to}_m \tau_1 \to \tau_2 \qquad \Gamma \vdash \tau_1 \Rightarrow e_2 \leadsto {}^{\Rightarrow}\check{e}}{\Gamma \vdash e_1 \triangleright e_2 \leadsto \left( {}^{\Box}_{\Rightarrow}{}_{\checkmark} \dot{\check{e}} {}^{\tau}_{\Rightarrow} \right) \triangleleft_m {}^{\Rightarrow}\check{e} {}^{\tau_2}_{\Rightarrow}}$$

$$\boxed{\Gamma \vdash \tau \Rightarrow e \leadsto \check{e}^{\Rightarrow}}$$

**MarkSubsume**

$$\frac{e \text{ subsumable} \qquad \Gamma \vdash e \leadsto \dot{\check{e}} {}^{\tau_2}_{\Rightarrow} \qquad \tau_1 \sim_m \tau_2}{\Gamma \vdash \tau_1 \Rightarrow e \leadsto {}^{\tau_1}_{\Rightarrow}{}_m \dot{\check{e}} {}^{\tau_2}_{\Rightarrow}}$$

**MarkAnaFun**

$$\frac{\tau \blacktriangleright^{\to}_{m_1} \tau_2 \to \tau_3 \qquad \tau_1 \sim_{m_2} \tau_2 \qquad \Gamma,\, x? : \tau_1 \vdash \tau_3 \Rightarrow e \leadsto {}^{\Rightarrow}\check{e}}{\Gamma \vdash \tau \Rightarrow \lambda x? : \tau_1 \mapsto e \leadsto {}^{\tau}_{\Rightarrow} \lambda x? : \tau_1 \mapsto_{m_1, m_2} {}^{\Rightarrow}\check{e} {}^{\Box}_{\Rightarrow}}$$

$$\boxed{e \leadsto \check{p}}$$

**MarkProgram**

$$\frac{\varnothing \vdash e \leadsto \check{e}^{\Rightarrow}}{e \leadsto {}^{\Box}_{\Rightarrow}{}_{\checkmark} \check{e}^{\Rightarrow}}$$

Fig. 2. Marking (from scratch)

The MarkVar rule marks a variable expression with a synthesized type found by looking up the variable in the context. Unlike in MLC, the context lookup judgment $x_m : \tau \in \Gamma$, defined in Figure 9, is a total function of $x$ and $\Gamma$, returning $m$ and $\tau$. If $x$ is an entry in $\Gamma$ then $m = \checkmark$ because the $x$ is not free, and the associated $\tau$ is returned; if not, then $m = \times$ and $\tau = ?$.

The MarkAsc rule marks a type ascription by analytically marking the expression body $e$ against the ascribed type $\tau$, resulting in ${}^{\Rightarrow}\check{e}$, and synthesizing $\tau$ as the type of the whole expression.

The MarkSynFun rule synthesizes a type for a function abstraction by first synthetically marking the body in the extended context. Note the extension of the context with a binder $x?$, which is either ? or some $x$. In the former case, since a binding hole does not bind any variables, we define $\Gamma,\ ? : \tau = \Gamma$. In the latter case, we interpret $\Gamma,\ x : \tau$ as simply extending an association list. The marked result in MarkSynFun is a marked abstraction with both marks set to $\checkmark$, because these mark positions indicate errors related to the analyzed type of the abstraction, and this is the synthetic rule. For the same reason, the body of the abstraction is analyzed against $\Box$, that is, it is not analyzed against any type. Finally, the abstraction synthesizes a function type with its domain determined by the annotation and its codomain determined by the body's synthesized type.

MarkAp reflects the standard bidirectional typing rule for function applications, with the added step of the *matched arrow judgment* from gradual type theory, defined in Figure 9. The expression

$$\frac{}{x_\times : \,? \,\in\, \varnothing}$$

$$\boxed{x_m : \tau \,\in\, \Gamma}$$

$$\frac{}{x_\checkmark : \tau \,\in\, \Gamma,\, x : \tau} \qquad \frac{x \neq x' \qquad x_m : \tau \,\in\, \Gamma}{x_m : \tau \,\in\, \Gamma,\, x' : \tau'}$$

$$\boxed{\tau \blacktriangleright_m^{\rightarrow} \tau \to \tau}$$

$$\frac{}{? \blacktriangleright_\checkmark^{\rightarrow} \,? \to \,?} \qquad \frac{}{\tau_1 \to \tau_2 \blacktriangleright_\checkmark^{\rightarrow} \tau_1 \to \tau_2} \qquad \frac{\tau \neq \,? \qquad \mathsf{hd}(\tau) \neq \,\rightarrow}{\tau \blacktriangleright_\times^{\rightarrow} \,? \to \,?}$$

$$\boxed{\tau \sim_m \tau}$$

$$\frac{}{? \sim_\checkmark \tau} \qquad \frac{}{\tau \sim_\checkmark \,?}$$

$$\frac{\tau_1 \sim_{m_1} \tau_3 \qquad \tau_2 \sim_{m_2} \tau_4 \qquad m_1 \sqcap m_2 = m}{\tau_1 \to \tau_2 \sim_m \tau_3 \to \tau_4} \qquad \frac{\tau_1 \neq \,? \qquad \tau_2 \neq \,? \qquad \mathsf{hd}(\tau_1) \neq \mathsf{hd}(\tau_2)}{\tau_1 \sim_\times \tau_2}$$

Fig. 3. Total Side Conditions

in function position is marked in synthetic mode, obtaining synthesized type $\tau$. Then $\tau$ is matched against the arrow type to obtain $\tau_1 \to \tau_2$ and mark $m$. This matching judgment is a necessary step for two independent reasons: because the unknown type ? is convertible to $? \to ?$, and because side conditions in this version of the calculus are total functions that return marks indicating success. The unknown type and an arrow type match the arrow form successfully, returning mark $\checkmark$, while any other type fails to match, returning the error mark $\times$ and unknown types in their error recovering capacity. The minimal type system shown in the figure only has the unknown and arrow types, so the error case is actually impossible to reach, but this would no longer be true if base types, product types, etc. were present, as they are in the accompanying Agda mechanization.

Now we proceed to the analytic marking rules. Most syntactic forms do not have special typing behavior when typed in analytic mode; they simply ignore the analyzed type, mark in synthetic mode, and compare the analyzed and synthesized types. Such forms are called *subsumable*, and include all forms in this minimal language except function abstractions. The analytic marking rule for these forms, Mark Subsume, first marks the expression in synthetic mode, then checks the consistency of the analyzed and synthesized types. The consistency judgment checks whether two types match, up to the replacement of some subterms with the unknown type. The rules are given in Figure 9. The rule for arrow types uses the expression $m_1 \sqcap m_2$, which is defined by the property that $m_1 \sqcap m_2 = \checkmark$ if and only if $m_1 = \checkmark$ and $m_2 = \checkmark$. Our minimal type system has the curious property that *all* pairs of types are consistent. Again, this would not be true if other type forms were added. The only output of the consistency judgment is the mark. In the case of Mark Subsume, this mark is placed on the marked analytic expression form.

The Mark Ana Fun rule specifies how abstractions are marked in analytic mode. The analyzed type is matched against the arrow type, with the resulting mark placed on the abstraction, because it is a type error to find a function abstraction where a non-function was expected. The resulting domain type is checked for consistency with the abstraction's annotation, resulting in the second mark placed on the abstraction, because the expected and found input types must match. Finally, the body is marked in analytic mode against the codomain of the analyzed type.

A bare expression is marked as a program by marking it in synthetic mode in the empty context, since the top level of a program has no expected type.

The marking operation from bare expressions to marked programs induces a notion of *well-markedness* on marked programs. Intuitively, a marked program is well-marked if it is the result of

$$
\begin{array}{rcll}
c & \in & \text{Child} & ::= \quad \text{One} \mid \text{Two} \\
\alpha & \in & \text{SimpleAction} & ::= \quad \text{InsertVar}(x) \mid \text{WrapFun} \mid \text{WrapAp}(c) \mid \text{WrapAsc} \\
& & & \mid \quad \text{Delete} \mid \text{Unwrap}(c) \mid \text{SetAnn}(\tau) \mid \text{SetAsc}(\tau) \\
& & & \mid \quad \text{InsertBinder}(x?) \mid \text{DeleteBinder}
\end{array}
$$

Fig. 4. Actions

marking its underlying bare expression correctly. To formalize this, we utilize a *marking erasure* function, denoted $\diamond \check{p} = e$. Erasure recursively removes marks and synthesized and analyzed types, resulting in the underlying bare expression. Well-markedness is then defined as such:

*Definition 3.1 (Well-Markedness).* A marked program $\check{p}$ is *well-marked* if $\diamond \check{p} \rightsquigarrow \check{p}$.

Well-markedness is used to express the correctness of Incremental MALC in Section 5.

### 3.3 Edit Actions

Let us now consider how users edit terms in the language. This is formalized in an *action calculus*, consisting of a set of simple actions $\alpha$, representing the kinds of edits the user can make, a set of localized actions $A$, consisting of simple actions paired with a location in the program, and an action performance judgment $e_1 \xrightarrow{A} e_2$, which performs a localized action in a bare expression.

Figure 11 defines the syntax of simple edit actions. They consist of actions to insert constructors at a leaf ($\text{InsertVar}(x)$), actions to insert constructors wrapped around a subterm (WrapFun, $\text{WrapAp}(c)$, WrapAsc), an action to delete a subterm (Delete), an action to unwrap a constructor from a subterm ($\text{Unwrap}(c)$), actions to change types in the surface syntax ($\text{SetAnn}(\tau)$, $\text{SetAsc}(\tau)$), and actions to insert or delete binders ($\text{InsertBinder}(x?)$, DeleteBinder). Wrapping or unwrapping a constructor with multiple children requires specifying which child position to wrap or unwrap around, hence the child argument to these actions, $c$. We do not model fine grained type edits in the formalism, since in a simple type system these are independent of the incremental type maintenance behavior, instead abstracting them into the wholesale type edit actions $\text{SetAnn}(\tau)$ and $\text{SetAsc}(\tau)$.

A localized action $A$ is a simple action $\alpha$ paired with a path into the expression from the root, represented as a list of child elements. Not every such path identifies a valid subexpression.

The action performance judgment $e_1 \xrightarrow{A} e_2$ states that the localized action $A$, when performed on the bare expression $e_1$, results in the new bare expression $e_2$. The rules for this judgment are straightforward, matching the intuitive meaning of the actions. The formal definition can be found in the Agda mechanization. While the choice of action language is an input to the theory, we are able to validate our choice in one respect by proving the completeness of our set of actions, in the sense that any program structure can be reached from any other by some sequence of actions. This is formalized by the following theorem:

THEOREM 3.2 (ACTION COMPLETENESS). *For any bare expressions $e_1$ and $e_2$, there exists a sequence of localized actions $\overline{A}$ such that $e_1 \xrightarrow{\overline{A}} e_2$ (where $e_1 \xrightarrow{\overline{A}} e_2$ denotes iterated action performance).*

## 4 Incremental MALC By Example

Before formally defining Incremental MALC, let us develop some intuition by following a small example edit trace. In Incremental MALC, programs are represented as incremental expressions, which mirror the marked expressions of the previous section but attach to each type a dirty bit, either *dirty*, $\star$, or *clean*, $\bullet$, which mediates the propagation of type information updates. The beginning of each subsection below displays the edit in a syntax closer to what the user would see, omitting type information except for occurrences of ×. Note that only the names of *simple* edit actions are displayed, with the location of the edit implied by its effect. For reference, the arrows representing steps in this section are tagged with names of the corresponding inference rules in section 5.

Term (1) displays the initial state of the program, which begins as an empty expression hole synthesizing the unknown type (?) and not analyzed against a type (□). Since there is no inconsistency between the analyzed and synthesized types, there is no consistency error on this expression. The types are furthermore annotated with • symbols, indicating that no type information needs to be propagated through the program at this time.

$$\overset{\square^\bullet}{\Rightarrow}_\checkmark ? \overset{?^\bullet}{\Rightarrow} \tag{1}$$

## 4.1 Inserting a Variable

$$\boxed{? \xrightarrow{\text{InsertVar}(x)} x_\times}$$

Suppose the user's first action is to insert an identifier $x$ into the empty hole. This is modeled as the performance of the action $\text{InsertVar}(x)$, localized at the hole. The immediate result is shown in term (2).

$$\xrightarrow{\text{InsertVar}(x)} \quad \overset{\square^\star}{\Rightarrow}_\checkmark x_\times \overset{?^\star}{\Rightarrow} \tag{2}$$

The variable $x$ is free at this location in the program, so it is given an error mark ×. Additionally, the analyzed and synthesized types are dirtied, meaning they have been added to the update propagation frontier. The analyzed type, though unchanged, is dirtied because the expression being analyzed is new. The synthesized type of the new variable is set to the variable's type. As it happens in this case, since the variable is free, it again synthesizes the unknown type.

At this point the editor, perhaps while waiting for the user's next action, would take two steps to propagate this new type information. The update propagation dynamics is nondeterministic, allowing either of the two dirtied types to step first. There is, however, a preferred prioritization order that minimizes redundant computation in the general case, as discussed in Section 6.3. In this running example, update propagation steps are taken in this order.

First the dirty analyzed type takes a step, bringing us to term (3).

$$\xrightarrow{\text{(NewAna)}} \quad \overset{\square^\bullet}{\Rightarrow}_\checkmark x_\times \overset{?^\star}{\Rightarrow} \tag{3}$$

The analyzed type encounters a subsumable syntactic form, so it updates the consistency mark, which in this case remains ✓. The analyzed □• is cleaned since its immediate ramifications have been calculated; it is no longer on the frontier.

The synthesized type, being at the root of the entire program, has no downstream information to propagate to, so it simply exits the frontier, bringing us to term 4, which has no dirty types. Such a term is called *quiescent*.

$$\xrightarrow{\text{(TopStep)}} \quad \overset{\square^\bullet}{\Rightarrow}_\checkmark x_\times \overset{?^\bullet}{\Rightarrow} \tag{4}$$

## 4.2 Wrapping a Function Application

$$\boxed{x_\times \xrightarrow{\text{WrapAp}(\text{One}),\ \text{InsertNum}(1)} x_\times \triangleleft 1}$$

Next, the user wraps the current expression in a function application form, then inserts a number literal into the argument position before update propagation. The result is shown in term (5). Note that base types are assumed for this example, and included in the mechanization.

$$\xrightarrow{\text{WrapAp(One) ... InsertNum(1)}} \quad \overset{\square\star}{\underset{\checkmark}{\Rightarrow}} \left( \overset{\square\star}{\underset{\checkmark}{\Rightarrow}} x_\times \overset{?\star}{\underset{\times}{\Rightarrow}} \right) \vartriangleleft_\checkmark \left( \overset{\square\star}{\underset{\checkmark}{\Rightarrow}} 1 \overset{\text{num}\star}{\Rightarrow} \right) \overset{\square\star}{\Rightarrow} \quad (5)$$

The leftmost two $\square^\star$ propagate similarly to the step from term (2) to term (3), resulting in term (6).

$$\xmapsto{(\text{NewAna})} \xmapsto{(\text{NewAna})} \quad \overset{\square\bullet}{\underset{\checkmark}{\Rightarrow}} \left( \overset{\square\bullet}{\underset{\checkmark}{\Rightarrow}} x_\times \overset{?\bullet}{\Rightarrow} \right) \vartriangleleft_\checkmark \left( \overset{\square\star}{\underset{\checkmark}{\Rightarrow}} 1 \overset{\text{num}\star}{\Rightarrow} \right) \overset{\square\star}{\Rightarrow} \quad (6)$$

Next the synthesized type from the function position of the application propagates outwards. The unknown type is matched against the function type as $? \to ?$. The argument expression is analyzed against the domain type $?$, and the entire application form synthesizes the codomain type $?$. This new analytic and synthetic type data is dirtied. The application's error mark remains $\checkmark$, because matched function type succeeded. The result is term (7).

$$\xmapsto{(\text{NewSynAp})} \quad \overset{\square\bullet}{\underset{\checkmark}{\Rightarrow}} \left( \overset{\square\bullet}{\underset{\times}{\Rightarrow}} x_\times \overset{?\bullet}{\Rightarrow} \right) \vartriangleleft_\checkmark \left( \overset{?\bullet}{\Rightarrow} 1 \overset{\text{num}\star}{\Rightarrow} \right) \overset{?\bullet}{\Rightarrow} \quad (7)$$

The newly analyzed $?^\star$ against the subsumable number literal in the argument position is then propagated, setting the consistency mark to $\checkmark$ again because the analyzed $?$ is consistent with the synthesized num. The synthesized $?^\star$ exits the frontier, having reached the root of the program. These two steps result in the quiescent program shown in term (8).

$$\xmapsto{(\text{NewAna})} \xmapsto{(\text{NewSyn})} \xmapsto{(\text{TopStep})} \quad \overset{\square\bullet}{\underset{\checkmark}{\Rightarrow}} \left( \overset{\square\bullet}{\underset{\checkmark}{\Rightarrow}} x_\times \overset{?\bullet}{\Rightarrow} \right) \vartriangleleft_\checkmark \left( \overset{?\bullet}{\Rightarrow} 1 \overset{\text{num}\bullet}{\Rightarrow} \right) \overset{?\bullet}{\Rightarrow} \quad (8)$$

## 4.3 Wrapping a Function Abstraction

$$\boxed{x_\times \vartriangleleft 1 \xrightarrow{\text{WrapFun, SetAnn(bool}\to\text{num), InsertBinder}(x)} \lambda x : (\text{bool} \to \text{num}) \mapsto (x \vartriangleleft_\times 1)}$$

Now the user wraps the program in a function abstraction, resulting in term (9). The binding position of the abstraction as well as its type annotation are initially empty holes, and a few types have been added to the update propagation frontier.

$$\xrightarrow{\text{WrapFun}} \quad \overset{\square\star}{\underset{\checkmark}{\Rightarrow}} \lambda? : ?^\bullet \mapsto_{\checkmark,\checkmark} \left( \overset{\square\star}{\underset{\checkmark}{\Rightarrow}} \left( \overset{\square\bullet}{\underset{\checkmark}{\Rightarrow}} x_\times \overset{?\bullet}{\Rightarrow} \right) \vartriangleleft_\checkmark \left( \overset{?\bullet}{\Rightarrow} 1 \overset{\text{num}\bullet}{\Rightarrow} \right) \overset{?\bullet}{\Rightarrow} \right) \overset{\square\star}{\Rightarrow} \quad (9)$$

Before update propagation, suppose the annotation is edited to bool $\to$ num. This new type in the surface syntax is placed in the update propagation frontier.

$$\xrightarrow{\text{SetAnn(bool}\to\text{num)}} \quad \overset{\square\star}{\underset{\checkmark}{\Rightarrow}} \lambda? : (\text{bool} \to \text{num})^\star \mapsto_{\checkmark,\checkmark} \left( \overset{\square\star}{\underset{\checkmark}{\Rightarrow}} \left( \overset{\square\bullet}{\underset{\checkmark}{\Rightarrow}} x_\times \overset{?\bullet}{\Rightarrow} \right) \vartriangleleft_\checkmark \left( \overset{?\bullet}{\Rightarrow} 1 \overset{\text{num}\bullet}{\Rightarrow} \right) \overset{?\bullet}{\Rightarrow} \right) \overset{\square\star}{\Rightarrow} \quad (10)$$

Still before any update propagation steps, let the user insert the variable $x$ in the binding position of the abstraction. This demonstrates an important property of the calculus: bindings, being nonlocal connections in the analysis of the program, are not updated via update propagation steps. They are updated atomically as part of the edit action that affects them. In this case, as soon as a binding for $x$ is inserted around the occurrence of $x$, the occurrence's error mark is set to $\checkmark$, since it is no longer free, and it synthesizes the type dictated by the binding annotation.

$$\xrightarrow{\text{InsertBinder}(x)} \quad \overset{\square\star}{\underset{\checkmark}{\Rightarrow}} \lambda x : (\text{bool} \to \text{num})^\star \mapsto_{\checkmark,\checkmark} \left( \overset{\square\star}{\underset{\checkmark}{\Rightarrow}} \left( \overset{\square\bullet}{\underset{\checkmark}{\Rightarrow}} x_\checkmark \overset{(\text{bool}\to\text{num})^\star}{\Rightarrow} \right) \vartriangleleft_\checkmark \left( \overset{?\bullet}{\Rightarrow} 1 \overset{\text{num}\bullet}{\Rightarrow} \right) \overset{?\bullet}{\Rightarrow} \right) \overset{\square\star}{\Rightarrow} \quad (11)$$

It is a challenge to efficiently maintain type information in response to binding changes such as the one above; the most obvious strategies require traversing the body of a binding construct or traversing the program's spine above a variable occurrence, but these operations may incur costs linear in the size of the program. It is this task to which we apply the order maintenance data structure, as described in detail in Section 6. Formally, however, binding calculations such as updating the bound $x$ above are assumed to occur atomically during action performance.

Now let us consider the propagation of updates in the current program state. The first two steps propagate the analyzed type and the annotation on the abstraction, resulting in a new analyzed type for the body and a new synthesized type for the abstraction. Function abstractions, being non-subsumable, use the expected type to find the expected type of the body, and only synthesize a type if not analyzed against a type. The result of these first two steps is shown in term (12).

$$\xLongrightarrow[\Rightarrow]{\square^{\bullet}} \lambda x : (\text{bool} \to \text{num})^{\bullet} \mapsto_{\checkmark,\checkmark} \overset{\longmapsto(\textsc{NewAnnFun})\longmapsto(\textsc{NewAnaFun})}{\left( \xLongrightarrow[\Rightarrow]{\square^{\bigstar}} \left( \xLongrightarrow[\Rightarrow]{\square^{\bullet}} x_{\checkmark} \xLongrightarrow[\Rightarrow]{(\text{bool}\to\text{num})^{\bigstar}} \right) \lhd_{\checkmark} \left( \xLongrightarrow[\Rightarrow]{?^{\bullet}} 1 \xLongrightarrow[\Rightarrow]{\text{num}^{\bullet}} \right) \xLongrightarrow[\Rightarrow]{?^{\bigstar}} \right)} ((\text{bool}\to\text{num})\to?)^{\bigstar}$$
(12)

Next the analyzed type on the variable is checked against the synthesized type, and then the synthesized type propagates through the function application. The result of these two steps is shown in term (13).

$$\xLongrightarrow[\checkmark]{\square^{\bullet}} \lambda x : (\text{bool} \to \text{num})^{\bullet} \mapsto_{\checkmark,\checkmark} \overset{\longmapsto(\textsc{NewAna})\longmapsto(\textsc{NewSynAp})}{\left( \xLongrightarrow[\checkmark]{\square^{\bullet}} \left( \xLongrightarrow[\checkmark]{\square^{\bullet}} x_{\checkmark} \xLongrightarrow[\Rightarrow]{(\text{bool}\to\text{num})^{\bullet}} \right) \lhd_{\checkmark} \left( \xLongrightarrow[\Rightarrow]{\text{bool}^{\bigstar}} 1 \xLongrightarrow[\Rightarrow]{\text{num}^{\bullet}} \right) \xLongrightarrow[\Rightarrow]{\text{num}^{\bigstar}} \right)} ((\text{bool}\to\text{num})\to?)^{\bigstar}$$
(13)

The analyzed type on the argument is compared with the synthesized type, and the synthesized type of the body propagates to a new synthesized type for the abstraction. The result is term (14).

$$\xLongrightarrow[\checkmark]{\square^{\bullet}} \lambda x : (\text{bool} \to \text{num})^{\bullet} \mapsto_{\checkmark,\checkmark} \overset{\longmapsto(\textsc{NewAna})\longmapsto(\textsc{NewSynFun})}{\left( \xLongrightarrow[\checkmark]{\square^{\bullet}} \left( \xLongrightarrow[\checkmark]{\square^{\bullet}} x_{\checkmark} \xLongrightarrow[\Rightarrow]{(\text{bool}\to\text{num})^{\bullet}} \right) \lhd_{\checkmark} \left( \xLongrightarrow[\times]{\text{bool}^{\bullet}} 1 \xLongrightarrow[\Rightarrow]{\text{num}^{\bullet}} \right) \xLongrightarrow[\Rightarrow]{\text{num}^{\bullet}} \right)} ((\text{bool}\to\text{num})\to\text{num})^{\bigstar}$$
(14)

Finally, the synthesized type of the whole program exits the update propagation frontier, having reached the root. The result is the quiescent program shown in term (15).

$$\xLongrightarrow[\checkmark]{\square^{\bullet}} \lambda x : (\text{bool} \to \text{num})^{\bullet} \mapsto_{\checkmark,\checkmark} \overset{\longmapsto(\textsc{TopStep})}{\left( \xLongrightarrow[\checkmark]{\square^{\bullet}} \left( \xLongrightarrow[\checkmark]{\square^{\bullet}} x_{\checkmark} \xLongrightarrow[\Rightarrow]{(\text{bool}\to\text{num})^{\bullet}} \right) \lhd_{\checkmark} \left( \xLongrightarrow[\times]{\text{bool}^{\bullet}} 1 \xLongrightarrow[\Rightarrow]{\text{num}^{\bullet}} \right) \xLongrightarrow[\Rightarrow]{\text{num}^{\bullet}} \right)} ((\text{bool}\to\text{num})\to\text{num})^{\bullet}$$
(15)

This example, though small, illustrates the fundamental ideas behind Incremental MALC. Edit actions locally update annotations and place them in the propagation frontier, and a small-step dynamics propagates these changes through the rest of the program in a series of local steps. In this syntax these updates propagate from left to right, an idea which has been formalized to prove termination for the dynamics. Once a program is quiescent and no more steps are possible, its marks and type annotations will be correct with respect to the non-incremental MALC.

## 5 Incremental MALC

We now formally specify Incremental MALC, introduced by example in the previous section.

### 5.1 Incremental Program Syntax

Incremental MALC operates on *incremental programs*, the syntax for which is defined in Figure 5.

Incremental programs are the same as the marked programs of Section 3, except that the types within, both those appearing in the surface syntax and stored analyzed or synthesized types, are

$$\overline{A} \quad ::= \quad \cdot \mid A, \overline{A}$$

$$\frac{p_1 \xrightarrow{A} p_2 \quad p_2 \xmapsto{\overline{A}} p_3}{p_1 \xmapsto{A,\overline{A}} p_3} \qquad \frac{p_1 \longmapsto p_2 \quad p_2 \xmapsto{\overline{A}} p_3}{p_1 \xmapsto{\overline{A}} p_3} \qquad \frac{\neg \exists p'. \; p \longmapsto p'}{p \xmapsto{} p}$$

| $\circ$ | $\in$ | DirtyBit | $::=$ | $\star \mid \bullet$ |
|---|---|---|---|---|
| $e^{\Rightarrow}$ | $\in$ | SynExp | $::=$ | $\dot{e}^{\sigma^{\circ}}$ |
| $\dot{e}$ | $\in$ | ConExp | $::=$ | $? \mid x_m \mid \lambda x? : \tau^{\circ} \mapsto_{m,m} {}^{\Rightarrow}e \mid {}^{\Rightarrow}e \vartriangleleft_m {}^{\Rightarrow}e \mid {}^{\Rightarrow}e : \tau^{\circ}$ |
| ${}^{\Rightarrow}e$ | $\in$ | AnaExp | $::=$ | ${}^{\sigma^{\circ}}_{\Rightarrow m}e^{\Rightarrow}$ |
| $p$ | $\in$ | Program $\subseteq$ AnaExp | $::=$ | ${}^{\square^{\circ}}_{\Rightarrow \checkmark}e^{\Rightarrow}$ |

Fig. 5. Incremental Syntax

Fig. 6. Interleaved action and update propagation

additionally annotated with a dirty bit, either *dirty*, $\star$, or *clean*, $\bullet$. Well-markedness is defined on incremental programs by coercing them to marked programs by erasing the dirty bits.

Types marked as dirty with $\star$ are members of the update propagation frontier, and represent a location where an update step is possible. Usually $\sigma^\star$ indicates that $\sigma$ has recently been set to a possibly new value, and the immediate ramifications of this value have not been computed yet. However, this is not always the case. For example, an analyzed type is annotated with $\star$ when the expression beneath it changes, rather than when the type itself changes.

### 5.2 Central Judgments and Properties

These are the two central judgment forms of the calculus. The action performance judgment, written $p \xrightarrow{A} p'$, applies a localized edit action $A$ to the program $p$, resulting in new program $p'$. The update propagation judgment, written $p \longmapsto p'$, describes the small step propagation dynamics that cause updates to static information throughout the program.

The judgment $p \xmapsto{\overline{A}} p'$, defined in Figure 6, combines action performance and update propagation to model the top-level behavior of the system. In the editor, propagations are run automatically and concurrently with action performance until none are possible. The combined judgment applies all actions in sequence, interleaved nondeterministically with arbitrary update propagation steps, until no actions remain and no steps are possible.

The primary correctness properties of this system are the following, which guide our definitions. Some additional intuition about the proofs are included in the appendix, and the full proofs are mechanized in Agda, discussed further in Section 5.5:

Validity ensures that the incremental analysis produces correct results with respect to the original, non-incremental theory. Here, well-formedness is an invariant on annotated programs that is preserved by actions and update steps. It is formally specified in the appendix. Informally, a program is well-formed if its type information is locally correct according to the typing rules of MALC, except possibly on the update propagation frontier.

THEOREM 5.1 (VALIDITY). *If program $p$ is well-formed and $p \xmapsto{\overline{A}} p'$, then $p'$ is well-marked.*

Convergence guarantees that any valid update propagation steps may be executed at any time during editing, and in any order, including interleaved with user edit actions, and the same resulting annotated program will be obtained.

**Theorem 5.2 (Convergence).** *If program $p$ is well-formed, $p \overset{\overline{A}}{\longmapsto} p_1$, and $p \overset{\overline{A}}{\longmapsto} p_2$, then $p_1 = p_2$.*

Termination guarantees that update propagation cannot continue forever, which guarantees the eventual correctness of the analysis provided a finite number of edit actions have been performed.

**Theorem 5.3 (Termination).** *There is no infinite sequence $\{p_n\}_{n=0}^{\infty}$ such that $\forall n.\ p_n \longmapsto p_{n+1}$.*

### 5.3 Action Performance

The action performance judgment for analytic expressions, $\Gamma \vdash \overset{\Rightarrow}{e_1} \overset{\alpha}{\to} \overset{\Rightarrow}{e_2}$, means that performing the simple action $\alpha$ directly on the analytic expression $\overset{\Rightarrow}{e_1}$, appearing in the typing context $\Gamma$, results in the new analytic expression $\overset{\Rightarrow}{e_2}$. It is defined by only one rule, which marks the analyzed type as new and applies the action to the synthetic expression within:

$$
\frac{\Gamma \vdash e_1^{\Rightarrow} \overset{\alpha}{\to} e_2^{\Rightarrow}}{\Gamma \vdash \overset{\sigma^\circ}{\underset{m}{\Rightarrow}} e_1^{\Rightarrow} \overset{\alpha}{\to} \overset{\sigma^\star}{\underset{m}{\Rightarrow}} e_2^{\Rightarrow}} \quad \text{\textsc{ActAna}}
$$

The corresponding judgment for synthetic expressions, $\Gamma \vdash e_1^{\Rightarrow} \overset{\alpha}{\to} e_2^{\Rightarrow}$, is where the core logic of action performance is specified. Each rule is designed to satisfy a few criteria. First, each action must change the structure of the program correctly (e.g. deletion should actually delete the subterm). This is formalized in the following lemma, which states that erasure maps action performance annotated expressions to action performance on bare expressions.

**Lemma 5.4 (Action Erasure).** *If $p \overset{A}{\to} p'$, then $\diamond p \overset{A}{\to} \diamond p'$.*

Secondly, each action must preserve the well-formedness invariant, the full definition of which can be found in the appendix. This allows us to prove the following essential lemma:

**Lemma 5.5 (Action Preservation).** *If $p$ is well-formed and $p \overset{\alpha}{\to} p'$, then $p'$ is well-formed.*

An additional criterion that is not essential for the formal properties of the system but is important for the practical instantiation of the system is that actions may add but never remove types from the update propagation frontier. Let us consider each rule in turn.

$$
\frac{x_m : \tau^\circ \in \Gamma}{\Gamma \vdash ? \overset{\sigma^\circ}{\Rightarrow} \xrightarrow{\text{InsertVar}(x)} x_m \overset{\tau^\star}{\Rightarrow}} \quad \text{\textsc{ActInsertVar}}
$$

$$
\text{\textsc{ActWrapFun}} \qquad \Gamma \vdash \dot{e} \overset{\sigma^\circ}{\Rightarrow} \xrightarrow{\text{WrapFun}} \lambda?:?^\bullet \mapsto_{\checkmark,\checkmark} \left( \overset{\square^\star}{\underset{\checkmark}{\Rightarrow}} \dot{e} \overset{\sigma^\star}{\Rightarrow} \right) \overset{\square^\star}{\Rightarrow}
$$

The rule **ActInsertVar** inserts a variable expression form into a hole. Formally, the context is queried to find the variable's mark and synthesized type, just as in the non-incremental MALC. The synthesized type is added to the update propagation frontier. The rule **ActWrapFun** wraps a subexpression in a lambda abstraction, with the binder and annotation initialized as holes. The synthesized type of the form is added to the update propagation frontier even though update propagation flowing from the other $\square^\star$ will eventually overwrite it. If it were not added to the frontier, the program would settle into the same state, and the correctness properties of the system would not be endangered. It is added to satisfy the well-formedness invariant, which is defined in terms of local consistency between adjacent data in the syntax tree. In other words, it is added not to guarantee correctness but to support the proof of correctness.

$$\text{ActWrapApOne}$$

$$\Gamma \vdash \dot{e} \overset{\sigma^\circ}{\Rightarrow} \xrightarrow{\text{WrapAp(One)}} \left( \overset{\square^\star}{\Rightarrow}_\checkmark \dot{e} \overset{\sigma^\star}{\Rightarrow} \right) \triangleleft_\checkmark \left( \overset{\square^\star}{\Rightarrow}_\checkmark ? \overset{?^\star}{\Rightarrow} \right) \overset{\square^\star}{\Rightarrow}$$

The rule ActWrapApOne wraps a subexpression in the function position of a function application form. The argument position is filled with a hole. Following the ordinary marking rule, the function position is not analyzed against a type.

$$\text{ActWrapApTwo}$$

$$\Gamma \vdash \dot{e} \overset{\sigma^\circ}{\Rightarrow} \xrightarrow{\text{WrapAp(Two)}} \left( \overset{\square^\star}{\Rightarrow} ? \overset{?^\star}{\Rightarrow} \right) \triangleleft_\checkmark \left( \overset{?^\star}{\Rightarrow}_\checkmark \dot{e} \overset{\sigma^\star}{\Rightarrow} \right) \overset{?^\star}{\Rightarrow}$$

ActWrapApTwo is similar, wrapping the subexpression in the argument position instead. This rule "folds in" a step of update propagation by placing the argument's analyzed type and the application's synthesized type on the update propagation frontier, rather than just the function position's synthesized type.

$$\text{ActWrapAsc} \qquad\qquad\qquad \text{ActDelete}$$

$$\Gamma \vdash \dot{e} \overset{\sigma^\circ}{\Rightarrow} \xrightarrow{\text{WrapAsc}} \left( \overset{?^\star}{\Rightarrow}_\checkmark \dot{e} \overset{\sigma^\circ}{\Rightarrow} \right) : ?^\bullet \overset{?^\star}{\Rightarrow} \qquad\qquad \Gamma \vdash e \overset{\Rightarrow}{} \xrightarrow{\text{Delete}} ? \overset{?^\star}{\Rightarrow}$$

ActWrapAsc wraps the subexpression in a type hole ascription, and also folds in the propagation of this new ascription to the neighboring analyzed and synthesized positions. ActDelete simply replaces an expression with an expression hole synthesizing the unknown type. Note that each occurrence of $\circ$ in these rules is a metavariable for a dirty bit value. To reduce visual clutter, we assume that these metavariables implicitly share the subscript of the type they accompany.

$$\text{ActUnwrapFun}$$

$$\frac{x?_{m_4} : \tau_2^\circ \in \Gamma \qquad [\![ x?_{m_4} \overset{\tau_2}{\Rightarrow} / x? ]\!] e^{\Rightarrow} = \dot{e} \overset{\sigma_2^\circ}{\Rightarrow}}{\Gamma \vdash \lambda x? : \tau_1^\circ \mapsto_{m_1, m_2} \left( \overset{\sigma_1^\circ}{\Rightarrow}_{m_3} e^{\Rightarrow} \right) \overset{\sigma_2^\circ}{\Rightarrow} \xrightarrow{\text{Unwrap(One)}} \dot{e} \overset{\sigma_2^\star}{\Rightarrow}}$$

ActUnwrapFun unwraps a function abstraction from around its body. By deleting a binding location, this action has the potential to "unshadow" an outer binder for the same variable name, causing the occurrences of the variable that used to be bound to the deleted binder to be rebound to the outer binder. The first premise looks in the context for the type of the binder variable in the outer scope. The second premise uses a new judgment, the "variable update" judgment, defined in the appendix. Its notation evokes substitution because it sets the consistency mark and synthesized type of all free occurrences of $x?$ in $e^{\Rightarrow}$. The resulting synthetic expression, with the type marked as new, is the result of the action. Note the first premise is a variant of the context lookup judgment that accepts a binding argument, rather than just a variable. If $x? = ?$, this version of context lookup returns an arbitrary $m$ and $\tau$, and the variable update operation acts as the identity.

$$\text{ActUnwrapApOne} \qquad\qquad\qquad\qquad \text{ActUnwrapApTwo}$$

$$\Gamma \vdash \left( \overset{\sigma_1^\circ}{\Rightarrow}_{m_1} \dot{e} \overset{\sigma_2^\circ}{\Rightarrow} \right) \triangleleft_{m_2} \overset{\Rightarrow}{e} \overset{\sigma_3^\circ}{\Rightarrow} \xrightarrow{\text{Unwrap(One)}} \dot{e} \overset{\sigma_2^\star}{\Rightarrow} \qquad\qquad \Gamma \vdash \overset{\Rightarrow}{e} \triangleleft_{m_1} \left( \overset{\sigma_1^\circ}{\Rightarrow}_{m_2} \dot{e} \overset{\sigma_2^\circ}{\Rightarrow} \right) \overset{\sigma_3^\circ}{\Rightarrow} \xrightarrow{\text{Unwrap(Two)}} \dot{e} \overset{\sigma_2^\star}{\Rightarrow}$$

ActUnwrapAsc

$$\Gamma \vdash \left( \overset{\sigma_1^\circ}{\Rightarrow}_m \dot{e} \overset{\sigma_2^\circ}{\Rightarrow} \right) : \tau^\circ \overset{\sigma_1^\circ}{\Rightarrow} \xrightarrow{\text{Unwrap(One)}} \dot{e} \overset{\sigma_2^\star}{\Rightarrow}$$

Rules ActUnwrapApOne and ActUnwrapApTwo unwrap an application form from around its left or right child, deleting the other child expression. The synthesized type of expression is added to the update propagation frontier. ActUnwrapAsc is analogous, deleting the type ascription.

ActSetAnn

$$\Gamma \vdash \lambda x? : \tau_1^\circ \mapsto_{m_1,m_2} \overset{\sigma^\circ}{\Rightarrow} e \xrightarrow{\text{SetAnn}(\tau_2)} \lambda x? : \tau_2^\star \mapsto_{m_1,m_2} \overset{\sigma^\circ}{\Rightarrow} e$$

ActSetAsc

$$\Gamma \vdash \overset{\Rightarrow}{e} : \tau_1^\circ \overset{\sigma^\circ}{\Rightarrow} \xrightarrow{\text{SetAsc}(\tau_2)} \overset{\Rightarrow}{e} : \tau_2^\star \overset{\sigma^\circ}{\Rightarrow}$$

ActSetAnn and ActSetAsc update types in the surface syntax of the program, either in function annotations or type ascriptions. Each merely sets the type to the action's argument and dirties it.

ActInsertBinder

$$\frac{[\![ x_{\overset{\tau}{\checkmark}}/x ]\!] e^{\Rightarrow} \ = \ \dot{e} \overset{\sigma_3^\circ}{\Rightarrow}}{\Gamma \vdash \lambda? : \tau^\circ \mapsto_{m_1,m_2} \left( \overset{\sigma_1^\circ}{\Rightarrow}_{m_3} e^{\Rightarrow} \right) \overset{\sigma_2^\circ}{\Rightarrow} \xrightarrow{\text{InsertBinder}(x)} \lambda x : \tau^\circ \mapsto_{m_1,m_2} \left( \overset{\sigma_1^\circ}{\Rightarrow}_{m_3} \dot{e} \overset{\sigma_2^\star}{\Rightarrow} \right) \overset{\sigma_2^\circ}{\Rightarrow}}$$

ActInsertBinder inserts a variable name into the binder hole of a function abstraction. This captures all free occurrences of the variable in the body, which now must synthesize the annotated type and be marked as bound, as implemented in the variable update premise.

ActDeleteBinder

$$\frac{x?_{m_4} : \tau_2^\circ \in \Gamma \qquad [\![ x?_{m_4} \overset{\tau_2}{\Rightarrow}/x? ]\!] e^{\Rightarrow} \ = \ \dot{e} \overset{\sigma_3^\circ}{\Rightarrow}}{\Gamma \vdash \lambda x? : \tau_1^\circ \mapsto_{m_1,m_2} \left( \overset{\sigma_1^\circ}{\Rightarrow}_{m_3} e^{\Rightarrow} \right) \overset{\sigma_2^\circ}{\Rightarrow} \xrightarrow{\text{DeleteBinder}} \lambda? : \tau_1^\circ \mapsto_{m_1,m_2} \left( \overset{\sigma_1^\circ}{\Rightarrow}_{m_3} \dot{e} \overset{\sigma_2^\star}{\Rightarrow} \right) \overset{\sigma_2^\circ}{\Rightarrow}}$$

ActDeleteBinder deletes the current binder of a function abstraction. This rule is similar to ActUnwrapFun, with the same premises. The only difference is that it does not remove the abstraction form from the body, it just sets the binder to a hole. The reasoning is the same as for ActUnwrapFun, as deleting the binding variable results in a potential "unshadowing" of the variables in the body.

## 5.4 Update Propagation

Now we turn to update propagation steps. Like the action performance rules, each step is designed to satisfy certain criteria. First, steps cannot change the structure of the program. They are not edits, but automated calculations that trigger while the user is editing. This is formalized in the following lemma:

Lemma 5.6 (Update Step Erasure). *If $p \longmapsto p'$, then $\diamond p = \diamond p'$.*

Second, like actions, steps must maintain the well-formedness invariant:

Lemma 5.7 (Update Step Preservation). *If $p$ is well-formed and $p \longmapsto p'$, then $p'$ is well-formed.*

Third, to support the implementation, each step removes exactly one type from the update propagation frontier, but can add any number. Fourth, the steps must make progress towards termination, so that no infinite sequence of steps is possible. This manifests as each step progressing the frontier of new types along the "bidirectional information flow," which corresponds to the left-to-right order in the syntax.

At the top level, a program can either take an ordinary step as an analytic expression, or a special step only possible at the root. This TopStep rule allows a new type synthesized by the whole program to exit the update propagation frontier, having reached the root of the program.

$$
\frac{\text{InsideStep}}{\underset{\checkmark}{\overset{\Box_1^\circ}{\Rightarrow}} e_1^{\rightarrow} \mapsto \underset{\checkmark}{\overset{\Box_2^\circ}{\Rightarrow}} e_2^{\rightarrow}}{\underset{\checkmark}{\overset{\Box_1^\circ}{\Rightarrow}} e_1^{\rightarrow} \longmapsto \underset{\checkmark}{\overset{\Box_2^\circ}{\Rightarrow}} e_2^{\rightarrow}}
\qquad
\frac{\text{TopStep}}{\underset{\checkmark}{\overset{\sigma^\star}{\Rightarrow}} \dot{e} \overset{}{\Rightarrow} \longmapsto \overset{\Box^\circ}{\Rightarrow} \dot{e} \overset{\sigma^*}{\Rightarrow}}
$$

Now we turn our attention to steps for analytic and synthetic expressions. These judgments form a kind of "mutual congruence," meaning that an outer expression may step by stepping within a subexpression, regardless of whether the outer expression is analytic or synthetic and whether the subexpression is analytic or synthetic. Let us consider the rules for stepping directly. s

$$
\frac{\text{StepSyn}}{\tau \sim_{m_2} \sigma}{\overset{\tau^*}{\underset{m_1}{\Rightarrow}} \dot{e} \overset{\sigma^\star}{\Rightarrow}} \atop {\mapsto \overset{\tau^*}{\underset{m_2}{\Rightarrow}} \dot{e} \overset{\sigma^*}{\Rightarrow}}
\qquad
\frac{\text{StepAna}}{\dot{e} \text{ subsumable} \qquad \sigma_1 \sim_{m_2} \sigma_2}{\overset{\sigma_1^\star}{\underset{m_1}{\Rightarrow}} \dot{e} \overset{\sigma_2^\circ}{\Rightarrow}} \atop {\mapsto \overset{\sigma_1^*}{\underset{m_2}{\Rightarrow}} \dot{e} \overset{\sigma_2^\circ}{\Rightarrow}}
$$

The StepSyn rule handles a newly synthesized type under an analyzed type that is not □. When the term is actually being analyzed against a type, the synthesized type is only used to calculate the consistency mark, hence consistency is reevaluated and nothing is dirtied. The StepAna rule handles the alternative case in which consistency is recomputed, that being when the analyzed type is new. If the root of the expression $\dot{e}$ is not subsumable, then consistency should not be checked and this rule is not applicable. Otherwise, the analyzed and synthesized data are checked for consistency as in the previous rule.

$$
\frac{\text{StepAnaFun}}{\sigma_1 \blacktriangleright^{\rightarrow}_{m_5} \sigma_5 \rightarrow \sigma_6 \qquad \sigma_5 \sim_{m_6} \tau \qquad \text{FunSyn}(\sigma_1, \tau, \sigma_3) = \sigma_7}{\overset{\sigma_1^\star}{\underset{m_1}{\Rightarrow}} \lambda x? : \tau^\circ \mapsto_{m_2, m_3} \left( \overset{\sigma_2^\circ}{\underset{m_4}{\Rightarrow}} \dot{e} \overset{\sigma_4^\circ}{\Rightarrow} \right) \overset{\sigma_3^\circ}{\Rightarrow}} \atop {\mapsto \overset{\sigma_1^*}{\underset{\checkmark}{\Rightarrow}} \lambda x? : \tau^\circ \mapsto_{m_5, m_6} \left( \overset{\sigma_6^\star}{\underset{m_4}{\Rightarrow}} \dot{e} \overset{\sigma_3^\circ}{\Rightarrow} \right) \overset{\sigma_7^\star}{\Rightarrow}}
$$

StepAnaFun propagates a new analyzed type on a function abstraction. The first two premises correspond to those in the non-incremental MALC. The third premise uses a new function, FunSyn. This determines what type the abstraction should synthesize in terms of the analyzed type, the annotated type, and the type synthesized by the body. It is defined by the following equations:

$$
\begin{aligned}
\text{FunSyn}(\tau_1, \tau_2, \sigma) &= \Box \\
\text{FunSyn}(\Box, \tau_2, \Box) &= \Box \\
\text{FunSyn}(\Box, \tau_1, \tau_2) &= \tau_1 \rightarrow \tau_2
\end{aligned}
$$

If the abstraction is analyzed against some type, then it does not synthesize a type. If it is not analyzed against a type, then it synthesizes the appropriate function type between the annotated type and the body's type (unless the body does not synthesize a type). By including all three inputs, this function provides the correct behavior for abstractions in both analytic and synthetic mode.

STEPANNFUN

$$\dfrac{[\![x?\underset{\checkmark}{\overset{\tau}{\Rightarrow}}/x?]\!]e_1^{\Rightarrow} \;=\; e_2^{\Rightarrow}}{\overset{\sigma_1^{\circ}}{\underset{m_1}{\Rightarrow}}\lambda x? : \tau^{\star} \mapsto_{m_2,m_3} \left(\overset{\sigma_2^{\circ}}{\underset{m_4}{\Rightarrow}}e_1^{\Rightarrow}\right)\overset{\sigma_3^{\circ}}{\Rightarrow}}$$

$$\mapsto \overset{\sigma_1^{\star}}{\underset{m_1}{\Rightarrow}}\lambda x? : \tau^{\bullet} \mapsto_{m_2,m_3}\left(\overset{\sigma_2^{\circ}}{\underset{m_4}{\Rightarrow}}e_2^{\Rightarrow}\right)\overset{\sigma_3^{\circ}}{\Rightarrow}$$

STEPSYNFUN

$$\dfrac{\mathsf{FunSyn}(\sigma_1, \tau, \sigma_2) \;=\; \sigma_4}{\overset{\sigma_1^{\circ}}{\underset{m_1}{\Rightarrow}}\lambda x? : \tau^{\circ} \mapsto_{m_2,m_3}\left(\overset{\square^{\circ}}{\underset{m_4}{\Rightarrow}}\dot{e}\overset{\sigma_2^{\star}}{\Rightarrow}\right)\overset{\sigma_3^{\circ}}{\Rightarrow}}$$

$$\mapsto \overset{\sigma_1^{\circ}}{\underset{m_1}{\Rightarrow}}\lambda x? : \tau^{\circ} \mapsto_{m_2,m_3}\left(\overset{\square^{\circ}}{\underset{m_4}{\Rightarrow}}\dot{e}\overset{\sigma_2^{\star}}{\Rightarrow}\right)\overset{\sigma_4^{\star}}{\Rightarrow}$$

STEPANNFUN propagates a change in the annotated type of an abstraction. The new type causes a corresponding update in the synthesized types of the abstraction's bound variables, which is implemented by the variable update in the premise. The new annotation also affects the synthesized type and second mark of the abstraction. For convenience, this is handled by marking the analyzed type as new, allowing STEPANAFUN to trigger later and implement this logic, rather than repeating the premises in both rules. STEPSYNFUN propagates a newly synthesized type from the body of a function abstraction upwards by recomputing FunSyn on the new inputs.

STEPAP

$$\dfrac{\sigma_2 \blacktriangleright_{m_4}^{\rightarrow} \sigma_5 \rightarrow \sigma_6}{\left(\overset{\sigma_1^{\circ}}{\underset{m_1}{\Rightarrow}}\dot{e}\overset{\sigma_2^{\star}}{\Rightarrow}\right)\lhd_{m_2}\left(\overset{\sigma_3^{\circ}}{\underset{m_3}{\Rightarrow}}e^{\Rightarrow}\right)\overset{\sigma_4^{\circ}}{\Rightarrow}}$$

$$\mapsto \left(\overset{\sigma_1^{\circ}}{\underset{m_1}{\Rightarrow}}\dot{e}\overset{\sigma_2^{\star}}{\Rightarrow}\right)\lhd_{m_4}\left(\overset{\sigma_5^{\star}}{\underset{m_3}{\Rightarrow}}e^{\Rightarrow}\right)\overset{\sigma_6^{\star}}{\Rightarrow}$$

STEPASC

$$\dfrac{}{\left(\overset{\sigma_1^{\circ}}{\underset{m}{\Rightarrow}}\dot{e}\right) : \tau^{\star}\overset{\sigma_2^{\circ}}{\Rightarrow}}$$

$$\mapsto \left(\overset{\tau^{\star}}{\underset{m}{\Rightarrow}}\dot{e}\right) : \tau^{\bullet}\overset{\tau^{\star}}{\Rightarrow}$$

STEPAP propagates a newly synthesized type from the function position of an application, with the same matched arrow type premise as in MALC. The resulting analyzed type for the body and synthesized type for the application are added to the frontier. STEPASC propagates changes from an ascribed type, with the new type synthesized by the form and analyzed in the body.

## 5.5  Agda Mechanization

All definitions, lemmas, theorems, and proofs of Incremental MALC, including those referenced in this section, have been mechanically proven in the Agda proof assistant. One minor difference between the written theory and the mechanization is that the mechanization does not define the marked programs of Section 3, instead identifying them with the equivalent set of quiescent incremental programs. This choice frees us from needing two almost identical definitions of the central data structure of the theory. Termination is directly shown by proving that the inverse of the update propagation step relation is well-founded, which is then used to prove to the given form of the theorem.

## 5.6  Language Extensions

Although our presentation of Incremental MALC is intentionally given in terms of a minimal simply typed language, the ideas are applicable to a broader class of language features. The static semantics of each syntactic form in the MALC can be expressed as a function that determines each "output" of the node (the analyzed types of its children, its synthesized type, and its marks) in terms of the "inputs" of the node (its types in the surface syntax, its analyzed type, and the synthesized types of its children). These input and output positions are ordered according to the direction in which information can flow, with annotated types ordered by the left-to-right syntax order, types in the surface syntax being upstream of annotated types, and marks being downstream of annotated types.

This schema is informal, but it enabled us to easily extend the system with additional simply typed language features in the implementation and mechanization, including products and lists. For language features that require modifying the judgmental structure of the calculus, e.g. System F-style polymorphism, additional consideration would be needed to manage type substitution and edits to type bindings. For example, one could use singleton kinds rather than explicit substitution operations to track type variable identities [32]. For more general type-level computations, e.g. for dependent type systems, we need incremental term reduction—a problem for a future paper!

## 6 Implementation

The formalism of Incremental MALC primarily supports reasoning about the correctness properties of edit actions and update propagation steps. This section describes Malcom, our implementation of Incremental MALC. Critically, Malcom's edit actions and update propagation steps never need to traverse the unchanged portions of the program, including to build type contexts or search scopes for variable occurrences.

### 6.1 Ordered Mutable Trees

In Malcom, terms are represented as mutable trees with additional pointers. Cursors are represented by pointers to AST nodes, and each node stores a pointer to its parent, allowing constant time cursor movement up and down the tree. Nodes also store types and marks; both are mutable as well. Dirtiness is represented implicitly, by membership in a global priority queue, described later.

Moreover, terms in Malcom are decorated with two *timestamps* $(a, b)$, where $a$ is the node's position in a pre-order traversal of the full program and $b$ is the node's position in a post-order traversal of the full program. The two timestamps form an interval, with $a < b$, with the expected properties: a term's interval strictly contains all its children's intervals, and sibling intervals are disjoint and ordered. The reader can imagine these timestamps as recording the left-to-right order of MALC annotations. For binders, we will refer to these intervals as representing the binder's scope, since the intervals contain all terms below the binder.

These timestamps have two beneficial properties. First, term containment corresponds to interval containment, so Malcom can quickly test whether one term (for example, a binder) contains another (for example, a variable). Secondly, MALC update steps take place in timestamp order, with the analytic steps performed in pre-order and the synthetic steps performed in post-order, which makes timestamps useful for prioritizing update steps.

Timestamps are not simple integers; they are elements of a global order maintenance structure. An order maintenance structure is a finite, totally ordered collection of elements $e$ that allows fast creation and comparison of elements. Concretely, it supports just two basic operations:

- Insert($e$), which constructs a new element directly after $e$.
- Compare($e_1, e_2$), which compares the given elements in the total order.

Critically, both order maintenance operations are fast—O(1) with a small constant—and the order maintenance is preserved even as more elements are added. Malcom's implementation of order maintenance is based on Bender et al. [6] and extends the minimal API above with a variant of Insert that returns a new element directly before, instead of after, another.

### 6.2 Executing edit actions and update steps

Most update steps and edit actions, which simply add or remove tree nodes and update their marks, can be performed directly. However, operations that affect variables and binders are challenging. Edit actions like changing the type annotation on a binder, changing the binder name, or deleting a binder can affect the type of all variables bound by that binder, which can be arbitrarily far away.

*6.2.1 Variable and Binder Pointers*   Malcom thus adds additional pointers to the tree to track binding information. Each variables stores a pointer to its binding site (that is, to the abstraction term that introduces it), and each binding site stores an ordered set of pointers to its bound variables, represented by a splay tree ordered by pre-order time stamp. It is convenient to treat the root of the program as the "binding site" for free variables, so it also stores a separate set of free variables for each variable name.

These additional pointers make certain update steps faster. For example, edits to type annotations on binders (SetAnn($\tau$)) can now update all uses of the bound variable by simply iterating through the set of variable pointers at that binder. However, the pointer sets must also be maintained as edits occur.

Malcom also maintains a global map from variable names to the ordered set of binders for that variable name, whose importance will be explained shortly. These ordered sets are also represented by splay trees ordered by pre-order timestamp. Splay nodes also store both the post-order timestamp of each binder and also the maximum post-order timestamp in the subtree of the splay tree rooted at that splay node. The splay tree operations are implemented so as to maintain this additional piece of data. When binders are inserted or deleted, these ordered sets are directly updated.

*6.2.2 Edits to Variables*   First, consider edits to variables. When a variable is deleted, it need only be removed from its binding site's set of bound variables. Since the variable stores a pointer to this site, this can be done directly. However, when a variable expression is inserted, it is necessary to locate its binding site (or the root if it is free) and set up pointers between the variable and its binding site.

To do so, we first use the global variable-name-to-binder map to look up the set of all binders for the inserted variable name. Since binders can shadow each other, we need the lowest binder that contains the variable in question; this is the binder with the largest pre-order timestamp whose interval contains the variable's interval. The splay tree allows Malcom to find this node in $O(\log n)$ time; Malcom also splays this node to the root of the splay tree to make later operations using this binder faster.

*6.2.3 Edits to Binders*   Next, consider edits to binders. When a binder is created in the program, such as filling the empty binder of a function abstraction with a variable name, Malcom must check whether that binder shadows some outer binding and, if so, update all re-bound variables. Identifying the outer binding uses the same splay tree lookup as variable insertion. That binder has an ordered set of pointers to its bound variables; Malcom must determine which of these bound variables are now instead bound to the new binder.

The new binder is a descendant of the outer binder. Therefore its scope is some subinterval of the outer binder's scope. Conceptually, then, Malcom needs to split the outer binder's scope into three segments—an initial segment below only the outer binder, a middle segment below the new binder, and a final segment also only below the outer binder. To do so, it first considers the pre-order timestamp of the new binder, and splays that to the root of the outer binder's splay tree. The root and the left subtree now represent the initial segment. The right subtree is removed and split again, on the new binder's post-order timestamp; the right subtree of the result is now the final segment. The remaining middle segment now contains all bound variables of the new binder; it becomes the new binder's ordered set of bound variables. Then the initial and final segments are joined (an $O(\log n)$ operation on splay trees) and become the outer binder's new, smaller set of bound variables.

Deleting a binder is similar, but in reverse. The outer binder is found, and its split tree is split at the deleted binder's pre-order timestamp. The two halves become the initial and final segments, joined to the deleted binder's set of bound variables. The resulting larger set becomes the outer binder's new set of bound variables. In case of both insertion and deletion, the variable's global set of binders is also updated.

While complex, this splay-tree-based algorithm allows for algorithmically efficient handling of binders without the need to traverse the program to search for other binders or bound variables.

## 6.3 Prioritizing Update Steps

Another challenge is actually locating possible update steps. Malcom achieves this by maintaining a priority queue of all locations where update steps are possible (all dirtied types), represented as pointers into the AST. Each update step, then, simply involves popping a dirtied type location from the priority queue, performing the corresponding update step, and inserting any newly dirtied type locations into the priority queue. The action performance and update propagation step rules of Incremental MALC have been crafted to ensure that they can be implemented this way. For example, the premise to each update propagation step rule requires exactly one type to be dirty, and the conclusion cleans this type while dirtying some other types. In Malcom, this corresponds to popping an update location from the queue, mutating local information in the program, and pushing some new set of update locations onto the queue.

The priority queue is ordered using the term's timestamps. For new analyzed types or types in the surface syntax, the priority is the node's pre-order timestamp, since analysis steps are performed pre-order. For new synthesized types, it is the post-order timestamp. Since Incremental MALC is non-deterministic, the order *does not affect correctness*. However, the chosen order is more efficient in many cases. Since bidirectional type checking moves "left to right" through the program, and the chosen order is also a "left to right" order, Malcom will also push terms with larger timestamps than the ones it just popped. This means that Malcom tends to perform upstream updates before downstream updates, so as to avoid duplicate work.

Finally, there's one last edge case: term deletion. In the calculus, deletion is trivial because all data is local. However, with the global update queue, care must be taken when deleting subexpressions that contain update locations. When a subexpression is deleted, it is traversed, with each AST node within being marked as deleted. When an update location is popped form the queue, it is first checked that the location has not been deleted. If it has been, it is skipped, and popping continues. This is the only operation in Malcom that traverses an entire subexpression, but this cost is still proportional to the size of the edit and should be small for most reasonable editing patterns.

## 6.4 Unchanged Type Optimization

As defined in Incremental MALC, actions and update propagation steps dirty all types that could have potentially changed, even if they happen to remain the same. A simple optimization is to forgo dirtying types when unchanged. This cannot be applied in all cases, such as the dirtying of the analyzed type where an action is performed. In this cases, it is because the expression changes, not the type, that the analyzed type must be propagated. This optimization is, however, applicable to all update propagation steps, as they do not change the form of the expression.

This optimization is implemented in Malcom for all update propagation steps. However, this requires comparing dirtied types for structural equality, which is linear in the size of the type. This should not be an issue for typical programs, but could be addressed by using hash-consing on type syntax trees and comparing hashes.

## 6.5 Language Workbench

We have implemented Malcom, as described above, as a language workbench written in OCaml. It implements the data structures described above, including order maintenance, splay trees, variable and binder pointers, and the name-to-binder map. Traversal of the entire program is never necessary.

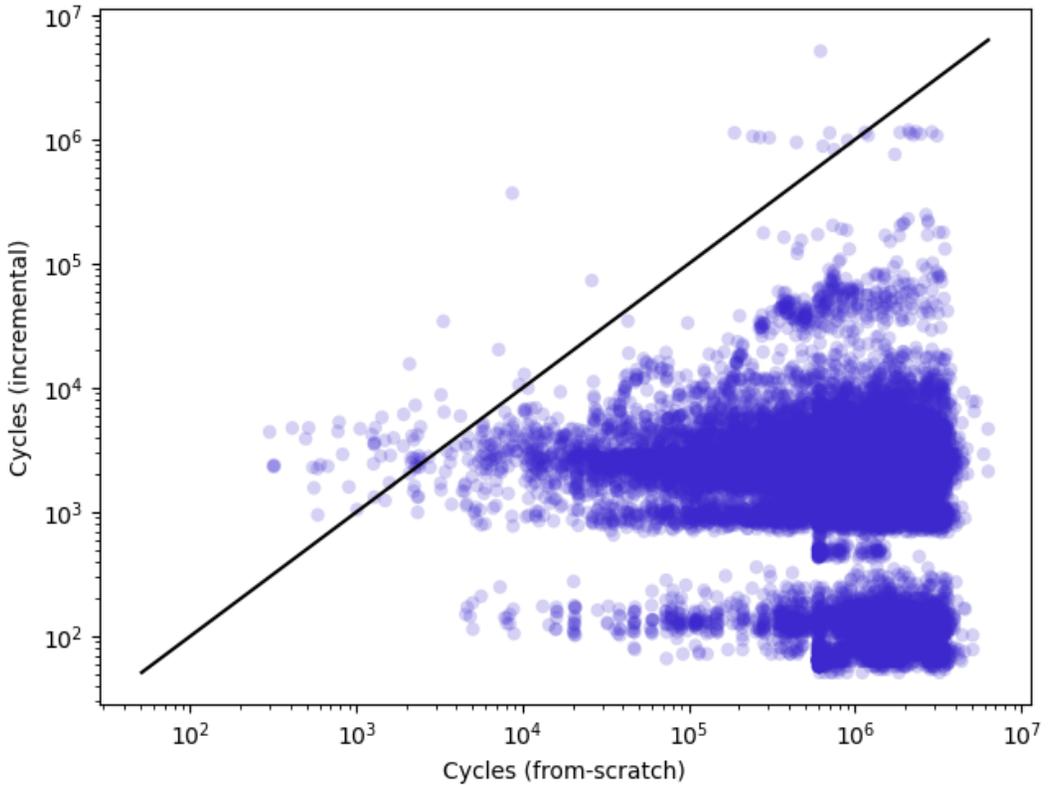

Fig. 7. Incremental vs. from-scratch editing times

We also wrote a simple debugging interface for Malcom that displays the program syntax tree in a textual format and renders it in a web browser. It maintains one cursor. There is a button for each edit action and each cursor movement. There is also an "update step" button to triggers the first update step and an "all update steps" button that triggers all update steps until the program is quiescent. Synthesized and analyzed typing information is hidden, except when they are members of the update propagation frontier, which allowed us to observe update propagation and debug our implementation. Note that the web interface is for debugging only, and while it updates type information incrementally, the actual interface is rendered from scratch at each step. We leave incremental rendering of the incrementally-computed type information for future work. We have extensively tested Malcom to ensure that the incremental output is always consistent with the from-scratch output.

## 7   Evaluation

In this section we present our evaluation of the efficiency of Malcom relative to from-scratch type checking.

### 7.1   Benchmark

We time Malcom on a synthesized edit trace constructed to test multiple aspects of Malcom. The first stage of the edit trace builds the program to a substantial size by constructing 100 nested copies of the merge sort algorithm. Each layer consists of a split function, a merge function, and a sort function. Each instance of the two helper functions, merge and split, is bound to a unique numbered identifier (`split_n` and `merge_n` for each layer n). Each sort function, on the other hand, is bound to the same identifier (`mergesort`). To capture long-distance binding patterns, each implementation of sort chooses which copy of split and which copy of merge to use uniformly at random from among those in scope. This tests two aspects of Malcom's handling of bindings.

(1) **Shadowing**. Each `mergesort` shadows the one before it, testing Malcom's ability to find ancestor binders efficiently.
(2) **Name Reuse**. The definitions of `split_n`, `merge_n`, and `mergesort` each use common local variable names, such as `x`. Thus, those variable names are bound in multiple non-overlapping scopes, and Malcom must bind each occurrence to the correct binder.

The edit trace builds this tower by inserting constructors at leaves of the program sketch until complete. At each point in construction, the children of the current node in the sketch are constructed in a random order, and this proceeds recursively. This randomness ensures that different patterns of variable interactions are tested: for example, a variable binding might be inserted into an abstraction after its body has been constructed, capturing the free occurrences in the body.

After the construction phase of the edit trace, random edits are applied throughout the program. Each of these edit sequences consists of moving to a uniformly random location in the program, applying a small change based on the node, and reverting the change. This is repeated 500 times. The changes include:

(1) Deleting a leaf node and inserting another in its place.
(2) Replacing a binder with another.
(3) Wrapping a constructor around a subterm.
(4) Alternatively, for constructors with one child subexpression (for which unwrapping does not delete subexpressions), unwrapping the constructor.

These edits test deletion, insertion, wrapping, and unwrapping, including updating types and binders. They also can introduce errors, demonstrating Malcom's incremental error marking ability.

## 7.2 Results

For each non-move edit, we measure the time it takes to perform the action and propagate updates until quiescent. Timing is done with the instruction rdtsc. The same edit is performed on a bare syntax tree represented with zipper data structure, and it is type checked from scratch according to the ordinary MALC marking procedure. The total time to edit and type check is compared between the incremental and from-scratch analyses. Note that action performance can be slower for Malcom than for the ordinary zipper representation, since Malcom computes changes to bindings at edit time. Figure 7 show a scatter plot comparing these times (in cycles) for Malcom and the from-scratch analysis. Each data point is one non-move edit to the program. Points above the diagonal (shown in black) are edits where the from-scratch analysis is faster than the incremental one, and below the diagonal is the converse. Almost all points are below the diagonal, meaning that incrementality provides a speedup in almost all cases. Many points are far below the diagonal, with incrementality providing one or more orders of magnitude of speedup. The data points are multi-modal, forming multiple horizontal bands. This indicates a fundamental algorithmic speedup: while the from-scratch analysis grows linearly in program size, empirically the incremental analysis shows hardly any growth. The horizontal bands are likely made up of different kinds of actions. The top most band displays a slight positive trend, indicating cases where the incremental analysis does grow with the program size. This band likely contains edits that affect binding structure.

In total time across all of these edits, Malcom provides whopping 275.96×speedup over from-scratch marking, demonstrating its substantial advantage over non-incremental analysis in providing continuous static feedback.

## 8  Related Work

There is a large body of work on incremental computation [18, 29], ranging from general frameworks such as Self-Adjusting Computation (SAC) [2], Adaption [14, 15], and incremental Datalog engines [36] to domain-specific approaches for lists [28], databases [20], and web browsers [13].

Order maintenance data structures, first introduced by [10], are a common data structure across these approaches, appearing in various forms in both general frameworks like SAC [3] and domain-specific systems like web browser layout engines [17].

When considering the specific problem tackled in this paper of incrementally maintaining type information in response to program edits, there have been prior efforts of both varieties as well. Our approach is a domain-specific approach targeted very specifically to bidirectional type checking with total error localization, building on recent advances, most notably the marked lambda calculus [37] which was described in Section 2. Fundamental to our approach is the use of a structure editor calculus to represent changes. We are most directly inspired by Hazelnut [23], which specified a bidirectionally typed structure editor calculus which was able to update type error marks locally at the cursor without needing recomputation. However, Hazelnut simply leaves edits that require distant changes to error marks undefined, making it impractical for real editing. Hazel, which was originally based on Hazelnut, now uses an approach based on the MLC, and we plan to integrate the approach from this paper into a future version of Hazel.

Szabó et al. [33] and Pacak et al. [25] take a more general approach by translating typing rules to a Datalog program whose database of derived semantic facts can be incrementally updated as the facts that encode the program are added and removed. This generic approach is worthy of continued investigation, but is limited by the generality of the incremental update algorithm for the chosen implementation of Datalog. One particular challenge is accounting for binding structure in a way that leverages the incrementality when encoded in Datalog. Pacak et al. [25] uses co-contextual typing [12], in which binding information is propagated bottom up rather than top down, to overcome issues related to binding structure updates. This approach, while fine grained enough to capture individual binding updates rather than whole-context changes, still requires traversing program spines when bindings are updated, potentially incurring linear time performance penalties.

Demers et al. [9] discusses incremental evaluation of attribute grammars, proposing two different approaches. As with Datalog, it is possible to encode many type systems as attribute grammars, and indeed the setting in this paper is the Cornell Program Synthesizer [34], one of the earliest structure editors and a pioneer in live semantic analysis. It may be that a generalized version of our approach would start to look like an incremental attribute grammar system, given analogies between synthesized and analyzed attributes and synthesized and analyzed types. We are not aware of a contemporary implementation of these ideas.

Our approach in this paper is distinctive relative to these related approaches in that we start by approaching the problem from type-theoretic first principles, developing a formal semantics for type information propagation and proving its metatheoretic properties, then implementing it in a highly specialized manner targeted to the problem of incremental bidirectional typing. We believe it will be fruitful to continue to compare generic approaches to our more direct approach, and hope that the novel benchmark task developed here will be ported to other systems and lead to productive competition (and even more realistic or punishing benchmarks).

In addition to approaches leveraging incrementality derived from edits, there are a number of approaches that rely on memoization or caching to improve the performance of language implementations. This includes a large body of work on incremental build systems [1, 22], which are in common use in industry. These systems generally operate at a coarse granularity, rebuilding entire compilation units (e.g. modules or packages) after changes. In contrast, our approach operates at a fine-grained level, on individual expressions. Consequently, our approach is more suitable for live programming environments where edits may occur many times per second.

This family of approaches also includes the system of Wachsmuth et al. [35], which develops a name resolution and type analysis engine using dependent "tasks", which each capture one step of analysis. This approach is language-independent but again less fine-grained than our approach, operating primarily at the file level in the implemented system. Some of the approaches we take in Malcom, namely the use of a priority queue to maintain the update propagation frontier, are reminiscent of task analysis engine based approaches.

Busi et al. [7] specify memoized versions of standard typing rules and implement this system to achieve speedups on various small functional and imperative benchmarks, but their approach sometimes requires traversing large parts of the program when a variable binding changes.

Aditya and Nikhil [4] present a system that incrementalizes the Hindley-Milner type inference algorithm based on a call-graph analysis. It specifically incrementalizes the unification phase of type checking, operating at the granularity of changes to individual declarations. Meertens [21] also presents an incremental system for unification-based type checking, which incrementally maintains the unification constraints generated by the program but does not incrementally solve them in general.

## 9 Discussion and Conclusion

In this work we provide a formal system and efficient implementation for maintaining type information in a simple editor calculus. To serve as the base theory, we introduce the marked and annotated lambda calculus (MALC), a novel variant of the marked lambda calculus.

Our incremental system is remarkably (pun intended) more efficient than from-scratch reanalysis. It contributes to the goal of live programming not just by being efficient, but by reducing the extent to which type checking blocks users from editing the program.

### 9.1 Future Work

Incremental MALC and Malcom present several opportunities for extension and improvement. One notable limitation of our approach is that although update propagation is less blocking than a batch analysis, it is still possible for some individual update propagation steps to take a long time. In particular, those that involve traversing every occurrence of a variable bound by a particular binder block editing for a time proportional to the number of bound variables. Although this would not seem to be much of a problem in practice, future work may refine the approach and address this possibility. One potential strategy is for the binder to track which of its variables have been updated with new information and update the variables one at a time during update propagation.

While Incremental MALC expresses highly incremental expression-level analysis, it does not provide an incremental solution for type-level computations. Consistency checks, for example, must be rerun in their entirety when only part of the input changes. Although the size of types might often remain modest compared with the size of the program, it may be valuable to extend Incremental MALC with incremental type-level operations in future work.

As described in Section 5.6, Incremental MALC as presented in this work fits within a more general schema that can accommodate additional language features. We leave it to future work to define, prove correct, and explore the expressivity of such a schema. We also leave to future work the engineering, and perhaps additional research, necessary to scale these ideas up to real-world type systems.

Incremental MALC supports liveness for only type-based editor services. To achieve large scale live programming requires incrementalized theories for other editor services, such as evaluation. There is much future work to be done in developing such theories.

## Data Availability Statement

This paper is accompanied by two artifacts: an Agda mechanization of the definitions and proofs for Incremental MALC (presented in Section 5) and an OCaml implementation and interactive workbench of the Malcom incremental type system (described in Section 6). The Agda mechanization can be found at https://github.com/hazelgrove/incremental-statics-agda and the workbench can be found at https://github.com/hazelgrove/incremental-hazelnut/tree/submission-march.

$$
\begin{array}{llll}
x & \in & \text{Variable} & \\
x? & \in & \text{Binding} & ::= \quad ?\mid x \\
\tau & \in & \text{Type} & ::= \quad ?\mid \tau\to\tau \\
e & \in & \text{BareExp} & ::= \quad ?\mid x\mid \lambda x?:\tau\mapsto e\mid e\triangleright e\mid e:\tau \\
m & \in & \text{Mark} & ::= \quad \checkmark\mid \times \\
\sigma & \in & \text{TypeOpt} & ::= \quad \Box\mid\tau \\
\check{e}^{\Rightarrow} & \in & \text{MarkedSynExp} & ::= \quad \dot{\check{e}}^{\sigma}_{\Rightarrow} \\
\check{e} & \in & \text{MarkedConExp} & ::= \quad ?\mid x_m\mid \lambda x?:\tau\mapsto_{m,m}{}^{\Rightarrow}\check{e}\mid {}^{\Rightarrow}\check{e}\triangleleft_m{}^{\Rightarrow}\check{e}\mid {}^{\Rightarrow}\check{e}:\tau \\
{}^{\Rightarrow}\check{e} & \in & \text{MarkedAnaExp} & ::= \quad {}^{\sigma}_{\Rightarrow_m}\check{e}^{\Rightarrow} \\
\check{p} & \in & \text{MarkedProgram} \subseteq \text{MarkedAnaExp} & ::= \quad {}^{\Box}_{\Rightarrow_{\checkmark}}\check{e}^{\Rightarrow} \\
\circ & \in & \text{DirtyBit} & ::= \quad \star\mid \bullet \\
e^{\Rightarrow} & \in & \text{SynExp} & ::= \quad \dot{e}^{\sigma}_{\Rightarrow} \\
\dot{e} & \in & \text{ConExp} & ::= \quad ?\mid x_m\mid \lambda x?:\tau^{\circ}\mapsto_{m,m}{}^{\Rightarrow}e\mid {}^{\Rightarrow}e\triangleleft_m{}^{\Rightarrow}e\mid {}^{\Rightarrow}e:\tau^{\circ} \\
{}^{\Rightarrow}e & \in & \text{AnaExp} & ::= \quad {}^{\sigma}_{\Rightarrow_m}e^{\Rightarrow} \\
p & \in & \text{Program} \subseteq \text{AnaExp} & ::= \quad {}^{\Box}_{\Rightarrow_{\checkmark}}e^{\Rightarrow}
\end{array}
$$

Fig. 8. Syntax of MALC and Incremental MALC

Fig. 9. Side condition judgments

## A  Definitions for Incremental MALC

The syntax of sorts relevant to MALC and Incremental MALC are given in Figure 8. The side conditions used for typing are defined in Figure 9. Note that these operations have been extended to accommodate optional types, as required in the action performance and update propagation step rules. The marking judgments for MALC are defined in Figure 10. Simple and localized actions are defined in Figure 11. The variable update judgment is defined in Figure 12. Figure 13 and Figure 14 define the rules for simple action performance and localized action performance. Figure 15 defines update propagation steps. Finally, Figure 16 defines the well-formedness invariant.

$$\boxed{\Gamma \vdash e \leadsto \check{e}^\Rightarrow}$$

**MarkHole**

$$\frac{}{\Gamma \vdash {?} \leadsto {?}\overset{?}{\Rightarrow}}$$

**MarkVar**

$$\frac{x_m : \tau \in \Gamma}{\Gamma \vdash x \leadsto x_m \overset{\tau}{\Rightarrow}}$$

**MarkAsc**

$$\frac{\Gamma \vdash \tau \Rightarrow e \leadsto \overset{\Rightarrow}{\check{e}}}{\Gamma \vdash e : \tau \leadsto \overset{\Rightarrow}{\check{e}} : \tau \overset{\tau}{\Rightarrow}}$$

**MarkSynFun**

$$\frac{\Gamma,\, x? : \tau_1 \vdash e \leadsto \dot{\check{e}} \overset{\tau_2}{\Rightarrow}}{\Gamma \vdash \lambda x? : \tau_1 \mapsto e \leadsto \lambda x? : \tau_1 \mapsto_{\checkmark,\checkmark} \left( \overset{\square}{\Rightarrow}_\checkmark \dot{\check{e}} \overset{\tau_2}{\Rightarrow} \right) \overset{\tau_1 \to \tau_2}{\Rightarrow}}$$

**MarkAp**

$$\frac{\Gamma \vdash e_1 \leadsto \dot{\check{e}} \overset{\tau}{\Rightarrow} \qquad \tau \overset{\to}{\blacktriangleright}_m \tau_1 \to \tau_2 \qquad \Gamma \vdash \tau_1 \Rightarrow e_2 \leadsto \overset{\Rightarrow}{\check{e}}}{\Gamma \vdash e_1 \triangleright e_2 \leadsto \left( \overset{\square}{\Rightarrow}_\checkmark \dot{\check{e}} \overset{\tau}{\Rightarrow} \right) \triangleleft_m \overset{\Rightarrow}{\check{e}} \overset{\tau_2}{\Rightarrow}}$$

$$\boxed{\Gamma \vdash \tau \Rightarrow e \leadsto \check{e}^\Rightarrow}$$

**MarkSubsume**

$$\frac{e \text{ subsumable} \qquad \Gamma \vdash e \leadsto \dot{\check{e}} \overset{\tau_2}{\Rightarrow} \qquad \tau_1 \sim_m \tau_2}{\Gamma \vdash \tau_1 \Rightarrow e \leadsto \overset{\tau_1}{\Rightarrow}_m \dot{\check{e}} \overset{\tau_2}{\Rightarrow}}$$

**MarkAnaFun**

$$\frac{\tau \overset{\to}{\blacktriangleright}_{m_1} \tau_2 \to \tau_3 \qquad \tau_1 \sim_{m_2} \tau_2 \qquad \Gamma,\, x? : \tau_1 \vdash \tau_3 \Rightarrow e \leadsto \overset{\Rightarrow}{\check{e}}}{\Gamma \vdash \tau \Rightarrow \lambda x? : \tau_1 \mapsto e \leadsto \overset{\tau}{\Rightarrow}_\checkmark \lambda x? : \tau_1 \mapsto_{m_1,m_2} \overset{\Rightarrow}{\check{e}} \overset{\square}{\Rightarrow}}$$

$$\boxed{e \leadsto \check{p}}$$

**MarkProgram**

$$\frac{\varnothing \vdash e \leadsto \check{e}^\Rightarrow}{e \leadsto \overset{\square}{\Rightarrow}_\checkmark \check{e}^\Rightarrow}$$

Fig. 10. Marking judgments

$$
\begin{array}{llll}
c & \in & \text{Child} & ::= \quad \text{One} \mid \text{Two} \\
\alpha & \in & \text{SimpleAction} & ::= \quad \text{InsertVar}(x) \mid \text{WrapFun} \mid \text{WrapAp}(c) \mid \text{WrapAsc} \\
& & & \mid \quad \text{Delete} \mid \text{Unwrap}(c) \mid \text{SetAnn}(\tau) \mid \text{SetAsc}(\tau) \\
& & & \mid \quad \text{InsertBinder}(x?) \mid \text{DeleteBinder} \\
\bar{s} & \in & \text{List[Sort]} & ::= \quad \cdot \mid s, \bar{s} \\
A & \in & \text{LocalizedAction} & ::= \quad (\alpha, \bar{c})
\end{array}
$$

Fig. 11. Actions

*Well-Formedness*   Well-formedness is defined in a syntax-directed way on all analytic expressions, with the WFSyn rule and the $\Gamma \vdash e^\Rightarrow$ judgment serving as a convenience for factoring the consistency check out of $\Gamma \vdash \overset{\Rightarrow}{e}$ for subsumable forms. The predicate utilizes the "directed consistency" relation, $a^\circ \succ a$. This relation holds when either the first argument is dirty or the arguments carry the same value. Programs are well-formed when this relation holds at every step of the bidirectional information flow. For example, in the rule WFAp, the analyzed type $\sigma_1^\circ$ is matched as an arrow type, resulting in a domain type, codomain type, and mark. This information, which was derived

$$\boxed{[\![x?_\tau \overset{m}{\Rightarrow}/x?]\!] \; e^{\Rightarrow} \; = \; e^{\Rightarrow}}$$

VarUpdateHole

$$\overline{[\![x_m \overset{\tau}{\Rightarrow}/x]\!] \; ?^{\sigma^\circ} \; = \; ?^{\sigma^\circ}}$$

VarUpdateVarEq

$$\overline{[\![x_{m_1} \overset{\tau}{\Rightarrow}/x]\!] \; x_{m_2} \overset{\sigma^\circ}{\Rightarrow} \; = \; x_{m_1} \overset{\tau^\star}{\Rightarrow}}$$

VarUpdateVarNeq

$$\frac{x \neq x'}{[\![x_{m_1} \overset{\tau}{\Rightarrow}/x]\!] \; x'_{m_2} \overset{\sigma^\circ}{\Rightarrow} \; = \; x'_{m_2} \overset{\sigma^\circ}{\Rightarrow}}$$

VarUpdateFunEq

$$\overline{[\![x_{m_1} \overset{\tau_1}{\Rightarrow}/x]\!] \left(\lambda x : \tau_2^\circ \mapsto_{m_2,m_3} \overset{\Rightarrow}{} e\right) \overset{\sigma^\circ}{\Rightarrow} \; = \; \left(\lambda x : \tau_2^\circ \mapsto_{m_2,m_3} \overset{\Rightarrow}{} e\right) \overset{\sigma^\circ}{\Rightarrow}}$$

VarUpdateFunNeq

$$\frac{x \neq x' \qquad [\![x_{m_1} \overset{\tau_1}{\Rightarrow}/x]\!] e_1^{\Rightarrow} \; = \; e_2^{\Rightarrow}}{[\![x_{m_1} \overset{\tau_1}{\Rightarrow}/x]\!] \left(\lambda x' : \tau_2^\circ \mapsto_{m_2,m_3} \overset{\sigma_1^\circ}{\underset{m_4}{\Rightarrow}} e_1^{\Rightarrow}\right) \overset{\sigma_2^\circ}{\Rightarrow} \; = \; \left(\lambda x' : \tau_2^\circ \mapsto_{m_2,m_3} \overset{\sigma_1^\circ}{\underset{m_4}{\Rightarrow}} e_2^{\Rightarrow}\right) \overset{\sigma_2^\circ}{\Rightarrow}}$$

VarUpdateAp

$$\frac{[\![x_{m_1} \overset{\tau}{\Rightarrow}/x]\!] e_1^{\Rightarrow} \; = \; e_3^{\Rightarrow} \qquad [\![x_{m_1} \overset{\tau}{\Rightarrow}/x]\!] e_2^{\Rightarrow} \; = \; e_4^{\Rightarrow}}{[\![x_{m_1} \overset{\tau}{\Rightarrow}/x]\!] \left(\overset{\sigma_1^\circ}{\underset{m_2}{\Rightarrow}} e_1^{\Rightarrow}\right) \lhd_{m_2} \left(\overset{\sigma_2^\circ}{\underset{m_3}{\Rightarrow}} e_2^{\Rightarrow}\right) \overset{\sigma_3^\circ}{\Rightarrow} \; = \; \left(\overset{\sigma_1^\circ}{\underset{m_2}{\Rightarrow}} e_3^{\Rightarrow}\right) \lhd_{m_2} \left(\overset{\sigma_2^\circ}{\underset{m_3}{\Rightarrow}} e_4^{\Rightarrow}\right) \overset{\sigma_3^\circ}{\Rightarrow}}$$

VarUpdateAsc

$$\frac{[\![x_{m_1} \overset{\tau_1}{\Rightarrow}/x]\!] e_1^{\Rightarrow} \; = \; e_2^{\Rightarrow}}{[\![x_{m_1} \overset{\tau_1}{\Rightarrow}/x]\!] \left(\overset{\sigma_1^\circ}{\underset{m_2}{\Rightarrow}} e_1^{\Rightarrow} : \tau_2^\circ\right) \overset{\sigma_2^\circ}{\Rightarrow} \; = \; \left(\overset{\sigma_1^\circ}{\underset{m_2}{\Rightarrow}} e_2^{\Rightarrow} : \tau_2^\circ\right) \overset{\sigma_2^\circ}{\Rightarrow}}$$

VarUpdateNone

$$\overline{[\![?_m \overset{\tau}{\Rightarrow}/?]\!] \; e^{\Rightarrow} \; = \; e^{\Rightarrow}}$$

Fig. 12. Variable update judgment

from the upstream type, are checked with the directed consistency relation against the downstream information found in the program. This check ensures that the information in the incremental program is locally consistent, except at the frontier of update propagation. Variants of the side condition judgments are introduced which operate on dirtied types, which produce dirty outputs if any inputs are dirty. Actions maintain this invariant by dirtying any information that may be newly inconsistent with its surroundings, and updates maintain the invariant by progressing the frontier according to the correct typing rules.

$$\boxed{\Gamma \vdash e^{\Rightarrow} \xrightarrow{\alpha} e^{\Rightarrow}}$$

**ACTINSERTVAR**
$$\frac{x_m : \tau^\circ \in \Gamma}{\Gamma \vdash ? \xRightarrow{\sigma^\circ} \xrightarrow{\mathsf{InsertVar}(x)} x_m \xRightarrow{\tau^\star}}$$

**ACTDELETE**
$$\frac{}{\Gamma \vdash e^{\Rightarrow} \xrightarrow{\mathsf{Delete}} ? \xRightarrow{?^\star}}$$

**ACTWRAPFUN**
$$\frac{}{\Gamma \vdash \dot{e} \xRightarrow{\sigma^\circ} \xrightarrow{\mathsf{WrapFun}} \lambda? : ?^\bullet \mapsto_{\checkmark,\checkmark} \left(\xRightarrow{\square^\star}_{\checkmark} \dot{e} \xRightarrow{\sigma^\star}\right) \xRightarrow{\square^\star}}$$

**ACTWRAPASC**
$$\frac{}{\Gamma \vdash \dot{e} \xRightarrow{\sigma^\circ} \xrightarrow{\mathsf{WrapAsc}} \left(\xRightarrow{?^\star}_{\checkmark} \dot{e} \xRightarrow{\sigma^\circ}\right) : ?^\bullet \xRightarrow{?^\star}}$$

**ACTWRAPAPONE**
$$\frac{}{\Gamma \vdash \dot{e} \xRightarrow{\sigma^\circ} \xrightarrow{\mathsf{WrapAp(One)}} \left(\xRightarrow{\square^\star}_{\checkmark} \dot{e} \xRightarrow{\sigma^\star}\right) \triangleleft_{\checkmark} \left(\xRightarrow{\square^\star}_{\checkmark} ? \xRightarrow{?^\star}\right) \xRightarrow{\square^\star}}$$

**ACTWRAPAPTWO**
$$\frac{}{\Gamma \vdash \dot{e} \xRightarrow{\sigma^\circ} \xrightarrow{\mathsf{WrapAp(Two)}} \left(\xRightarrow{\square^\star}_{\checkmark} ? \xRightarrow{?^\star}\right) \triangleleft_{\checkmark} \left(\xRightarrow{?^\star}_{\checkmark} \dot{e} \xRightarrow{\sigma^\star}\right) \xRightarrow{?^\star}}$$

**ACTUNWRAPFUN**
$$\frac{x?_{m_4} : \tau_2^\circ \in \Gamma \qquad [\![ x?_{m_4} \xRightarrow{\tau_2} /x? ]\!] e^{\Rightarrow} = \dot{e} \xRightarrow{\sigma_3^\circ}}{\Gamma \vdash \lambda x? : \tau_1^\circ \mapsto_{m_1,m_2} \left(\xRightarrow{\square^\circ}_{m_3} e^{\Rightarrow}\right) \xRightarrow{\sigma_2^\circ} \xrightarrow{\mathsf{Unwrap(One)}} \dot{e} \xRightarrow{\sigma_3^\star}}$$

**ACTUNWRAPASC**
$$\frac{}{\Gamma \vdash \left(\xRightarrow{\sigma_1^\circ}_m \dot{e} \xRightarrow{\sigma_2^\circ}\right) : \tau^\circ \xRightarrow{\sigma_3^\circ} \xrightarrow{\mathsf{Unwrap(One)}} \dot{e} \xRightarrow{\sigma_2^\star}}$$

**ACTUNWRAPAPONE**
$$\frac{}{\Gamma \vdash \left(\xRightarrow{\sigma_1^\circ}_{m_1} \dot{e} \xRightarrow{\sigma_2^\circ}\right) \triangleleft_{m_2} e^{\Rightarrow} \xrightarrow{\mathsf{Unwrap(One)}} \dot{e} \xRightarrow{\sigma_2^\star}}$$

**ACTUNWRAPAPTWO**
$$\frac{}{\Gamma \vdash {}^{\Rightarrow}e \triangleleft_{m_1} \left(\xRightarrow{\sigma_1^\circ}_{m_2} \dot{e} \xRightarrow{\sigma_2^\circ}\right) \xRightarrow{\sigma_3^\circ} \xrightarrow{\mathsf{Unwrap(Two)}} \dot{e} \xRightarrow{\sigma_2^\star}}$$

**ACTSETANN**
$$\frac{}{\Gamma \vdash \lambda x? : \tau_1^\circ \mapsto_{m_1,m_2} {}^{\Rightarrow}e \xRightarrow{\sigma^\circ} \xrightarrow{\mathsf{SetAnn}(\tau_2)} \lambda x? : \tau_2^\star \mapsto_{m_1,m_2} {}^{\Rightarrow}e \xRightarrow{\sigma^\circ}}$$

**ACTSETASC**
$$\frac{}{\Gamma \vdash {}^{\Rightarrow}e : \tau_1^\circ \xRightarrow{\sigma^\circ} \xrightarrow{\mathsf{SetAsc}(\tau_2)} {}^{\Rightarrow}e : \tau_2^\star \xRightarrow{\sigma^\circ}}$$

**ACTINSERTBINDER**
$$\frac{[\![ x_{\checkmark} \xRightarrow{\tau} /x ]\!] e^{\Rightarrow} = \dot{e} \xRightarrow{\sigma_3^\circ}}{\Gamma \vdash \lambda? : \tau^\circ \mapsto_{m_1,m_2} \left(\xRightarrow{\sigma_1^\circ}_{m_3} e^{\Rightarrow}\right) \xRightarrow{\sigma_2^\circ} \xrightarrow{\mathsf{InsertBinder}(x)} \lambda x : \tau^\circ \mapsto_{m_1,m_2} \left(\xRightarrow{\sigma_1^\circ}_{m_3} \dot{e} \xRightarrow{\sigma_3^\star}\right) \xRightarrow{\sigma_2^\circ}}$$

**ACTDELETEBINDER**
$$\frac{x?_{m_4} : \tau_2^\circ \in \Gamma \qquad [\![ x?_{m_4} \xRightarrow{\tau_2} /x? ]\!] e^{\Rightarrow} = \dot{e} \xRightarrow{\sigma_3^\circ}}{\Gamma \vdash \lambda x? : \tau_1^\circ \mapsto_{m_1,m_2} \left(\xRightarrow{\sigma_1^\circ}_{m_3} e^{\Rightarrow}\right) \xRightarrow{\sigma_2^\circ} \xrightarrow{\mathsf{DeleteBinder}} \lambda? : \tau_1^\circ \mapsto_{m_1,m_2} \left(\xRightarrow{\sigma_1^\circ}_{m_3} \dot{e} \xRightarrow{\sigma_3^\star}\right) \xRightarrow{\sigma_2^\circ}}$$

$$\boxed{\Gamma \vdash {}^{\Rightarrow}e \xrightarrow{\alpha} {}^{\Rightarrow}e}$$

**ACTANA**
$$\frac{\Gamma \vdash e_1^{\Rightarrow} \xrightarrow{\alpha} e_2^{\Rightarrow}}{\Gamma \vdash \xRightarrow{\sigma^\circ}_m e_1^{\Rightarrow} \xrightarrow{\alpha} \xRightarrow{\sigma^\star}_m e_2^{\Rightarrow}}$$

Fig. 13. Simple Action Performance

$$\boxed{\Gamma \vdash \dot{e} \xrightarrow{A} \dot{e}}$$

**ActFunRec**
$$\frac{x?, \tau^\circ : \Gamma \vdash {}^{\Rightarrow}e_1 \xrightarrow{(\alpha,\, \bar{c})} {}^{\Rightarrow}e_2}{\Gamma \vdash \lambda x? : \tau^\circ \mapsto_{m_1,m_2} {}^{\Rightarrow}e_1 \xrightarrow{(\alpha,\, (\mathsf{One},\, \bar{c}))} \lambda x? : \tau^\circ \mapsto_{m_1,m_2} {}^{\Rightarrow}e_2}$$

**ActAscRec**
$$\frac{\Gamma \vdash {}^{\Rightarrow}e_1 \xrightarrow{(\alpha,\, \bar{c})} {}^{\Rightarrow}e_2}{\Gamma \vdash {}^{\Rightarrow}e_1 : \tau^\circ \xrightarrow{(\alpha,\, (\mathsf{One},\, \bar{c}))} {}^{\Rightarrow}e_2 : \tau^\circ}$$

**ActApRecOne**
$$\frac{\Gamma \vdash {}^{\Rightarrow}e_1 \xrightarrow{(\alpha,\, \bar{c})} {}^{\Rightarrow}e_3}{\Gamma \vdash {}^{\Rightarrow}e_1 \triangleleft_m {}^{\Rightarrow}e_2 \xrightarrow{(\alpha,\, (\mathsf{One},\, \bar{c}))} {}^{\Rightarrow}e_3 \triangleleft_m {}^{\Rightarrow}e_2}$$

**ActApRecTwo**
$$\frac{\Gamma \vdash {}^{\Rightarrow}e_2 \xrightarrow{(\alpha,\, \bar{c})} {}^{\Rightarrow}e_3}{\Gamma \vdash {}^{\Rightarrow}e_1 \triangleleft_m {}^{\Rightarrow}e_2 \xrightarrow{(\alpha,\, (\mathsf{Two},\, \bar{c}))} {}^{\Rightarrow}e_1 \triangleleft_m {}^{\Rightarrow}e_2}$$

$$\boxed{\Gamma \vdash e^{\Rightarrow} \xrightarrow{A} e^{\Rightarrow}} \quad \boxed{\Gamma \vdash {}^{\Rightarrow}e \xrightarrow{A} {}^{\Rightarrow}e} \quad \boxed{p \xrightarrow{A} p}$$

**ActSynRec**
$$\frac{\Gamma \vdash \dot{e}_1 \xrightarrow{A} \dot{e}_2}{\Gamma \vdash \dot{e}_1 \overset{\sigma^\circ}{\Rightarrow} \xrightarrow{A} \dot{e}_2 \overset{\sigma^\circ}{\Rightarrow}}$$

**ActAnaRec**
$$\frac{\Gamma \vdash e_1^{\Rightarrow} \xrightarrow{A} e_2^{\Rightarrow}}{\Gamma \vdash \overset{\sigma^\circ}{\Rightarrow}_m e_1^{\Rightarrow} \xrightarrow{A} \overset{\sigma^\circ}{\Rightarrow}_m e_2^{\Rightarrow}}$$

**ActAnaLocal**
$$\frac{\Gamma \vdash {}^{\Rightarrow}e_1 \xrightarrow{\alpha} {}^{\Rightarrow}e_2}{\Gamma \vdash {}^{\Rightarrow}e_1 \xrightarrow{(\alpha,\, \cdot)} {}^{\Rightarrow}e_2}$$

**ActProgram**
$$\frac{\varnothing \vdash \overset{\Box_1^\circ}{\Rightarrow}_{\checkmark} e_1^{\Rightarrow} \xrightarrow{A} \overset{\Box_2^\circ}{\Rightarrow}_{\checkmark} e_2^{\Rightarrow}}{\overset{\Box_1^\circ}{\Rightarrow}_{\checkmark} e_1^{\Rightarrow} \xrightarrow{A} \overset{\Box_2^\circ}{\Rightarrow}_{\checkmark} e_2^{\Rightarrow}}$$

Fig. 14. Localized Action Performance

$$\boxed{\mathsf{FunSyn}(\sigma, \tau, \sigma) = \sigma}$$

$$
\begin{aligned}
\mathsf{FunSyn}(\tau_1, \tau_2, \sigma) &= \square \\
\mathsf{FunSyn}(\square, \tau_2, \square) &= \square \\
\mathsf{FunSyn}(\square, \tau_1, \tau_2) &= \tau_1 \to \tau_2
\end{aligned}
$$

$$\boxed{\stackrel{\Rightarrow}{e} \mapsto \stackrel{\Rightarrow}{e}}$$

**STEPSYN**

$$
\cfrac{\tau \sim_{m_2} \sigma}{
\stackrel{\tau^*}{\Rightarrow}_{m_1} \dot{e} \stackrel{\sigma^\star}{\Rightarrow}
}
$$

$$\mapsto \stackrel{\tau^*}{\Rightarrow}_{m_2} \dot{e} \stackrel{\sigma^*}{\Rightarrow}$$

**STEPANA**

$$
\cfrac{\dot{e}\ \text{subsumable} \qquad \sigma_1 \sim_{m_2} \sigma_2}{
\stackrel{\sigma_1^\star}{\Rightarrow}_{m_1} \dot{e} \stackrel{\sigma_2^\circ}{\Rightarrow}
}
$$

$$\mapsto \stackrel{\sigma_1^*}{\Rightarrow}_{m_2} \dot{e} \stackrel{\sigma_2^\circ}{\Rightarrow}$$

**STEPANNFUN**

$$
\cfrac{[\![ x?_\checkmark \stackrel{\tau}{/} x? ]\!] e_1^{\Rightarrow} = e_2^{\Rightarrow}}{
\stackrel{\sigma_1^\circ}{\Rightarrow}_{m_1} \lambda x? : \tau^\star \mapsto_{m_2, m_3} \left( \stackrel{\sigma_2^\circ}{\Rightarrow}_{m_4} e_1^{\Rightarrow} \right) \stackrel{\sigma_3^\circ}{\Rightarrow}
}
$$

$$\mapsto \stackrel{\sigma_1^\star}{\Rightarrow}_{m_1} \lambda x? : \tau^\bullet \mapsto_{m_2, m_3} \left( \stackrel{\sigma_2^\circ}{\Rightarrow}_{m_4} e_2^{\Rightarrow} \right) \stackrel{\sigma_3^\star}{\Rightarrow}$$

**STEPSYNFUN**

$$
\cfrac{\mathsf{FunSyn}(\sigma_1, \tau, \sigma_2) = \sigma_4}{
\stackrel{\sigma_1^\circ}{\Rightarrow}_{m_1} \lambda x? : \tau^\circ \mapsto_{m_2, m_3} \left( \stackrel{\sigma_2^\star}{\Rightarrow}_{m_4} \dot{e} \stackrel{\sigma_3^\circ}{\Rightarrow} \right) \stackrel{\sigma_3^\circ}{\Rightarrow}
}
$$

$$\mapsto \stackrel{\sigma_1^\circ}{\Rightarrow}_{m_1} \lambda x? : \tau^\circ \mapsto_{m_2, m_3} \left( \stackrel{\square^\circ}{\Rightarrow}_\checkmark \dot{e} \stackrel{\sigma_4^\star}{\Rightarrow} \right) \stackrel{\sigma_4^\star}{\Rightarrow}$$

**STEPANAFUN**

$$
\cfrac{\sigma_1 \blacktriangleright^{\to}_{m_5} \sigma_5 \to \sigma_6 \qquad \sigma_5 \sim_{m_6} \tau \qquad \mathsf{FunSyn}(\sigma_1, \tau, \sigma_3) = \sigma_7}{
\stackrel{\sigma_1^\star}{\Rightarrow}_{m_1} \lambda x? : \tau^\circ \mapsto_{m_2, m_3} \left( \stackrel{\sigma_2^\circ}{\Rightarrow}_{m_4} \dot{e} \stackrel{\sigma_3^\circ}{\Rightarrow} \right) \stackrel{\sigma_4^\circ}{\Rightarrow}
} \quad + \text{ cong. rules}
$$

$$\mapsto \stackrel{\sigma_1^*}{\Rightarrow}_\checkmark \lambda x? : \tau^\circ \mapsto_{m_5, m_6} \left( \stackrel{\sigma_6^\star}{\Rightarrow}_{m_4} \dot{e} \stackrel{\sigma_3^\circ}{\Rightarrow} \right) \stackrel{\sigma_7^\star}{\Rightarrow}$$

$$\boxed{e^{\Rightarrow} \mapsto e^{\Rightarrow}}$$

**STEPAP**

$$
\cfrac{\sigma_2 \blacktriangleright^{\to}_{m_4} \sigma_5 \to \sigma_6}{
\left( \stackrel{\sigma_1^\circ}{\Rightarrow}_{m_1} \dot{e} \stackrel{\sigma_2^\circ}{\Rightarrow} \right) \triangleleft_{m_2} \left( \stackrel{\sigma_3^\circ}{\Rightarrow}_{m_3} e^{\Rightarrow} \right) \stackrel{\sigma_1^\circ}{\Rightarrow}
}
$$

$$\mapsto \left( \stackrel{\sigma_1^\circ}{\Rightarrow}_{m_1} \dot{e} \stackrel{\sigma_2^\star}{\Rightarrow} \right) \triangleleft_{m_4} \left( \stackrel{\sigma_5^\star}{\Rightarrow}_{m_3} e^{\Rightarrow} \right) \stackrel{\sigma_6^\star}{\Rightarrow}$$

**STEPASC**

$$
\cfrac{}{
\left( \stackrel{\sigma_1^\circ}{\Rightarrow}_m \dot{e} \right) : \tau^\star \stackrel{\sigma_1^\circ}{\Rightarrow}
} \quad + \text{ cong. rules}
$$

$$\mapsto \left( \stackrel{\tau^\circ}{\Rightarrow}_m \dot{e} \right) : \tau^\bullet \stackrel{\tau^\star}{\Rightarrow}$$

$$\boxed{p \longmapsto p}$$

**INSIDESTEP**

$$
\cfrac{\stackrel{\square_1^\circ}{\Rightarrow}_\checkmark e_1^{\Rightarrow} \mapsto \stackrel{\square_2^\circ}{\Rightarrow}_\checkmark e_2^{\Rightarrow}}{
\stackrel{\square_1^\circ}{\Rightarrow}_\checkmark e_1^{\Rightarrow} \longmapsto \stackrel{\square_2^\circ}{\Rightarrow}_\checkmark e_2^{\Rightarrow}
}
$$

**TOPSTEP**

$$
\cfrac{}{
\stackrel{\square^\circ}{\Rightarrow}_\checkmark \dot{e} \stackrel{\sigma^\star}{\Rightarrow} \longmapsto \stackrel{\square^\circ}{\Rightarrow}_\checkmark \dot{e} \stackrel{\sigma^*}{\Rightarrow}
}
$$

Fig. 15. Update Propagation Steps

$$\boxed{\sigma^\circ \blacktriangleright_{m^\circ}^{\rightarrow} \sigma^\circ \rightarrow \sigma^\circ} \qquad \boxed{\sigma^\circ \sim_{m^\circ} \sigma^\circ}$$

$$\frac{\tau_1 \blacktriangleright_m^{\rightarrow} \tau_2 \rightarrow \tau_3}{\tau_1^\circ \blacktriangleright_{m^\circ}^{\rightarrow} \tau_2^\circ \rightarrow \tau_3^\circ} \qquad \frac{\tau_1 \sim_m \tau_2 \qquad \circ_1 \sqcup \circ_2 = \circ_3}{\tau_1^{\circ_1} \sim_{m^{\circ_3}} \tau_2^{\circ_2}}$$

$$\boxed{\Gamma \vdash e^{\Rightarrow}} \qquad \boxed{\Gamma \vdash {}^{\Rightarrow}e}$$

**WFHole**
$$\frac{?^{\bullet} \succ \sigma}{\Gamma \vdash ? \overset{\sigma^\circ}{\Rightarrow}}$$

**WFVar**
$$\frac{x_m : \sigma^\circ \in \Gamma \qquad \sigma^\circ \succ \tau}{\Gamma \vdash x_m \overset{\tau^\circ}{\Rightarrow}}$$

**WFAsc**
$$\frac{\tau_1^\circ \succ \tau_2 \qquad \tau_1^\circ \succ \tau_3 \qquad \Gamma \vdash \overset{\tau_2^\circ}{\Rightarrow}_m e^{\Rightarrow}}{\Gamma \vdash \tau_1^\circ : \overset{\tau_2^\circ}{\Rightarrow}_m e^{\Rightarrow} \overset{\tau_3^\circ}{\Rightarrow}}$$

**WFAp**
$$\frac{\sigma_1^\circ \blacktriangleright_{m_3^\circ}^{\rightarrow} \sigma_4^\circ \rightarrow \sigma_5^\circ \qquad \sigma_4^\circ \succ \sigma_2 \qquad \sigma_5^\circ \succ \sigma_3 \qquad m_3^\circ \succ m_1 \qquad \Gamma \vdash \overset{\Box^\circ}{\Rightarrow}_{\checkmark} \dot{e} \overset{\sigma_1^\circ}{\Rightarrow} \qquad \Gamma \vdash \overset{\sigma_2^\circ}{\Rightarrow}_{m_2} e^{\Rightarrow}}{\Gamma \vdash \left( \overset{\Box^\circ}{\Rightarrow}_{\checkmark} \dot{e} \right) \triangleleft_{m_1} \left( \overset{\sigma_2^\circ}{\Rightarrow}_{m_2} e^{\Rightarrow} \right) \overset{\sigma_3^\circ}{\Rightarrow}}$$

**WFSubsume**
$$\frac{\dot{e} \text{ subsumable} \qquad \sigma_1^\circ \sim_{m_2} \sigma_2^\circ \qquad m_2 \succ m_1 \qquad \Gamma \vdash \dot{e} \overset{\sigma_2^\circ}{\Rightarrow}}{\Gamma \vdash \overset{\sigma_1^\circ}{\Rightarrow}_{m_1} \dot{e} \overset{\sigma_2^\circ}{\Rightarrow}}$$

**WFFun**
$$\sigma_1^\circ \blacktriangleright_{m_5^\circ}^{\rightarrow} \sigma_5^\circ \rightarrow \sigma_6^\circ \qquad \sigma_5^\circ \sim_{m_6^\circ} \tau^\circ \qquad \sigma_6^\circ \succ \sigma_4 \qquad m_5^\circ \succ m_2$$

$$\frac{m_6^\circ \succ m_3 \qquad \mathsf{FunSyn}(\sigma_1^\circ, \tau^\circ, \sigma_3^\circ) \succ \sigma_4 \qquad \sigma_1^\circ \sim_{m_7^\circ} \sigma_4^\circ \qquad m_7^\circ \succ m_1 \qquad \Gamma, x? : \tau^\circ \vdash \overset{\sigma_2^\circ}{\Rightarrow}_{m_4} \dot{e} \overset{\sigma_3^\circ}{\Rightarrow}}{\Gamma \vdash \overset{\sigma_1^\circ}{\Rightarrow}_{m_1} \lambda x? : \tau^\circ \mapsto_{m_2, m_3} \left( \overset{\sigma_2^\circ}{\Rightarrow}_{m_4} \dot{e} \overset{\sigma_3^\circ}{\Rightarrow} \right) \overset{\sigma_4^\circ}{\Rightarrow}}$$

$$\boxed{a^\circ \succ a} \qquad \boxed{\vdash p}$$

**≻Dirty**
$$\frac{}{a_1^{\star} \succ a_2}$$

**≻Clean**
$$\frac{}{a^{\bullet} \succ a}$$

**WFProgram**
$$\frac{\varnothing \vdash p}{\vdash p}$$

Fig. 16. The well-formedness invariant

# B  Properties of Incremental MALC

This section describes the formal properties proven for Incremental MALC. The full proofs are available in the accompanying Agda mechanization.

We begin with validity, which is expressed in terms of the relation $p \xrightarrow{\overline{A}} p$, modeling interleaved action performance and update propagation, as well as the well-markedness condition on programs.

$$\frac{p_1 \xrightarrow{A} p_2 \quad p_2 \xrightarrow{\overline{A}} p_3}{p_1 \xrightarrow{A,\overline{A}} p_3} \qquad \frac{p_1 \longmapsto p_2 \quad p_2 \xrightarrow{\overline{A}} p_3}{p_1 \xrightarrow{\overline{A}} p_3} \qquad \frac{\neg \exists p'. \, p \longmapsto p'}{p \xmapsto{\,} p}$$

*Definition B.1 (Well-Markedness).*  A program $p$ is *well-marked* if $\diamond p \rightsquigarrow \check{p}$, where $\check{p}$ is equal to $p$ up to dirtiness of types.

THEOREM B.2 (VALIDITY).  *If program $p$ is well-formed and $p \xrightarrow{\overline{A}} p'$, then $p'$ is well-marked.*

Validity guarantees that the result of the incremental analysis agrees with the from-scratch static analysis. The following lemmas and definition are used in the proof:

LEMMA B.3 (ACTION PRESERVATION).  *If $p$ is well-formed and $p \xrightarrow{\alpha} p'$, then $p'$ is well-formed.*

LEMMA B.4 (UPDATE STEP PRESERVATION).  *If $p$ is well-formed and $p \longmapsto p'$, then $p'$ is well-formed.*

*Definition B.5 (Quiescent).*  An incremental program $p$ is *quiescent* if it contains no dirty types.

LEMMA B.6 (PROGRESS).  *If $p$ is well-formed, then it can take an update step if and only if it is not quiescent.*

LEMMA B.7 (QUIESCENT VALIDITY).  *If $p$ is well-formed and quiescent, then it is well-marked.*

These can be composed in a straightforward way to prove validity. Since $p$ is well-formed and $p \xrightarrow{\overline{A}} p'$, then by Lemma B.3 and Lemma B.4 (action preservation and update step preservation) $p'$ is also well-formed. No steps are possible from $p'$, Lemma B.6 (progress) ensures that it is quiescent. Lemma B.7 (quiescent validity) can then be applied to obtain the guarantee of well-markedness.

THEOREM B.8 (CONVERGENCE).  *If program $p$ is well-formed, $p \xrightarrow{\overline{A}} p_1$, and $p \xrightarrow{\overline{A}} p_2$, then $p_1 = p_2$.*

Convergence is proven with the help of the following lemmas:

LEMMA B.9 (ACTION ERASURE).  *If $p \xrightarrow{A} p'$, then $\diamond p \xrightarrow{A} \diamond p'$.*

LEMMA B.10 (UPDATE STEP ERASURE).  *If $p \longmapsto p'$, then $\diamond p = \diamond p'$.*

LEMMA B.11 (ACTION UNICITY).  *If $e \xrightarrow{A} e_1$ and $e \xrightarrow{A} e_2$, then $e_1 = e_2$.*

LEMMA B.12 (MARKING UNICITY).  *If $e \rightsquigarrow p_1$ and $e \rightsquigarrow p_2$, then $p_1 = p_2$.*

Assuming $p$ is well-formed, $p \xrightarrow{\overline{A}} p_1$, and $p \xrightarrow{\overline{A}} p_2$, then Lemma B.9 and Lemma B.10 (action erasure and update step erasure) imply that $\diamond p \xrightarrow{\overline{A}} \diamond p_1$ and $\diamond p \xrightarrow{\overline{A}} \diamond p_2$. By Lemma B.11 (action unicity), $\diamond p_1 = \diamond p_2$. By Theorem B.2 (validity), $p_1$ and $p_2$ are well-marked, meaning that $\diamond p_1 \rightsquigarrow p_1$ and $\diamond p_2 \rightsquigarrow p_2$. Using Lemma B.12 (marking unicity) and the fact that $\diamond p_1 = \diamond p_2$, we obtain $p_1 = p_2$.

THEOREM B.13 (TERMINATION).  *There is no infinite sequence $\{p_n\}_{n=0}^{\infty}$ such that $\forall n. \, p_n \longmapsto p_{n+1}$.*

Termination is proven by defining a well-founded ordering $<_P$ on programs such that if $p \longmapsto p'$, then $p' <_P p$. Informally, $p' <_P p$ if either $p'$ has strictly fewer dirty types in the surface syntax (that is, on a function annotation or type ascription), or $p'$ and $p$ have an equal number of such dirty types, but the update propagation frontier of $p'$ is further downstream than that of $p$ in the bidirectional flow. The syntax for incremental programs was designed to have the property that the left-to-right order in which annotations appear corresponds to the downstream order of information flow. All update rules except for those dealing with types in the surface syntax only dirty types to the right of the type they clean.